\documentclass[a4paper,12pt]{article}
\usepackage[round]{natbib}
\usepackage[english]{babel}
\usepackage{setspace}
\usepackage[utf8x]{inputenc}
\usepackage{graphicx} 
\usepackage[left=2cm,right=2cm,top=2cm,bottom=2cm]{geometry}
\usepackage{textcomp}
\usepackage[T1]{fontenc}
\usepackage{amssymb}
\usepackage{xcolor}
\usepackage{authblk}
\usepackage{ulem}
\usepackage{hyperref}
\usepackage{lineno}
\usepackage{mathabx}
\usepackage{appendix}
\usepackage{float}

\newcommand{\SO}{SO$_2$ }
\newcommand{\HO}{H$_2$O }

\title{
\large
\textbf{The Impact of Turbulent Vertical Mixing in the Venus Clouds on Chemical Tracers}
}

\author[1]{Maxence Lef{\`e}vre}
\author[2]{Emmanuel Marcq}
\author[2]{Franck Lef{\`e}vre}
\affil[1]{Department of Physics (Atmospheric, Oceanic and Planetary Physics), University of Oxford, Oxford, UK}
\affil[2]{LATMOS/IPSL, UVSQ Universit\'e Paris-Saclay, Sorbonne Universit\'e, CNRS, France}

\date{Accepted in Icarus}

\begin{document}

\maketitle
\section*{Abstract}
Venus’ clouds host a convective layer between roughly 50 and 60~km that mixes heat, momentum, and chemical species. Observations and numerical modelling have helped to understand the complexity of this region. However, the impact on chemistry is still not known. Here, we use for the first time a three-dimensional convection-resolving model with passive tracers to mimic \SO and \HO for two latitudinal cases. The tracers are relaxed towards a vertical profile in agreement with measured values, with a timescale varying over several orders of magnitude. The vertical mixing is quantified, it is strong for a relaxation timescale high in front of the convective timescale, around 4~hours. The spatial and temporal variability of the tracer due to the convective activity is estimated, with horizontal structures of several kilometres. At the Equator, the model is resolving a convective layer at the cloud top (70~km) suggested by some observations, the impact of such turbulent activity on chemical species is accounted for the first time. From the resolved convective plumes, a vertical eddy diffusion is estimated, consistent with past estimations from in-situ measurements, but several orders of magnitude higher than values used in 1D chemistry modelling. The results are compared to observations, with some spatial and temporal variability correlation, suggesting an impact of the convective layer on the chemical species.

\section{Introduction}

The strong dynamical activity inside the Venusian cloud layer has been assessed since the beginning of the Venus spacecraft exploration. Cloud images by the \textit{Mariner 10} mission \citep{Belt76b} and the \textit{Pioneer Venus} spacecraft \citep{Ross80} near the subsolar point showed cellular features suggesting convective cells with diameters between 200 and 1000~km. The convective activity was first measured by the \textit{Pioneer Venus} radio occultation experiment \citep{Seif80} from 50 to 55~km of altitude, and was then confirmed by the radio occultation experiment onboard the Magellan probe \citep{Hins95}. The \textit{VeGa} balloons flew in that altitude range close to the Equator and measured vertical winds between -4 and 2~m~s$^{-1}$ \citep{Link86a,Lore18} and convective cell diameter from several hundred meters to tens of kilometers \citep{Kerz86} around 54~km of altitude. The VeRa radio occultation device on board of \textit{Venus Express} studied in detail this convective layer and measured a strong latitudinal variability of the depth of the layer \citep{Tell09}, reaching 10~km close to 80$^{\circ}$ of latitude, almost twice the value of the equatorial regions. The radio occultation experiment in the \textit{Akatsuki} spacecraft measured variability of the convection depth with local time \citep{Imam17}, this layer being thicker a night.

In addition to the convection layer in the deep cloud layer, the Venus Monitoring Camera (VMC) observed the cellular features at the top of the cloud, about 70~km of altitude, at the subsolar point suggesting convective activity \citep{Mark07,Tito12}. A convective layer at this altitude is the main hypothesis for these observed structures, with measured convective cells from 20 to a few hundred of kilometres. However, the different radio occultation on board of \textit{Venus Express} and \textit{Akatsuki} radio occultation did not measure any clear neutral-stability layers at the subsolar point \citep{Ando18,Ando20}.

Gravity waves emitted from the convective layer have been observed by different space probes, \textit{Pioneer Venus} radio science observed evidenced small-scale waves with vertical wavelengths of about 7~km above and below the cloud layer \citep{Seif80,Coun80}, the \textit{Venus Express} instruments measured the wavelengths of the waves emitted above the cloud layer, ranging between about 2 and 3.5~km vertically \citep{Tell12} and from 2~km to hundreds of kilometres horizontally \citep{Pera08,Picc14}. The gravity waves in this region have also been studied with \textit{Akatsuki} \citep{Imam18,Mori21}.

Decades of spacecraft and ground-based observations of sulphur dioxide and water show highly variable abundance in the upper cloud deck, with timescales from hours to decades \citep{Marc20,Encr16,Vand17a,Vand17b}. Convection is one of the hypotheses for the short, from hours to days, term variability \citep{Marc13,Vand17a}. HST Imaging Spectrograph was used to observe Venus \citep{Jess15,Jess20} at cloud-top altitudes, albedo darkening was measured and explained by a possible increase of the convective vertical mixing and the injection of the unknown absorbing species. Using the SOIR/Venus Express CO$_2$ and CO profiles above the clouds and 1D photochemical model, \cite{Mahi21} estimated the vertical mixing from 80 to 140~km.

1D models have been developed to study the chemistry of the Venusian atmosphere. \cite{Kras07} and \cite{Kras13} focused on the lower atmosphere and \cite{Kras12}, \cite{Zhan12} and \cite{Park15} and \cite{Shao20} on the middle atmosphere. \cite{Yung09} and \cite{Bier20} and \cite{Rimm21} modelled the atmosphere from the surface to 110~km. These models cannot resolve the turbulent activity inside the cloud and therefore use the eddy diffusivity coefficient formalism to represent the different turbulent processes in the atmosphere like convection. There is a large uncertainty over the value of this coefficient in the Venusian atmosphere. These models showed that the mesospheric abundance of several species (especially \SO) was very sensitive to the eddy diffusivity values and vertical gradient.

Due to the lack of understanding of the turbulence inside the clouds of Venus, the  effect of the cloud convective layer and gravity waves on the chemistry and microphysics has not been studied in detail. Only \cite{Mcgo08} gave an insight into the change of optical depth due to the convection and gravity waves, using an idealized 2D (zonal/vertical) representation of the Venus cloud convective layer. \cite{More22} studied the vertical mixing due to the species stratification in the Venus lower atmosphere and clouds using Direct Numerical Simulation. High density-gradient magnitude regions are formed with increasing stratification and low stratification conditions produce a more uniform spatial distribution of the density.

To understand the turbulence activity inside the Venus cloud layer, the limited-area Venus mesoscale model (VMM) adapted from a terrestrial hydrodynamical solver \citep{Skam08} was developed at Laboratoire de M{\'e}t{\'e}orologie Dynamique by \citet{Lefe17} and later coupled to the full set of physical packages for Venus developed at Institut Pierre Simon Laplace (IPSL) \citep{Lebo10,Lebo16,Gara18} to simulate only a specified region of the planet with a fine resolution.

The 3D Large-Eddy Simulation (LES) mode was used to study the convection and small-scale gravity waves at different latitudes and local times \citep{Lefe17,Lefe18}. The resolved convection depth is consistent with observations with updrafts cells diameter of 20~km. The gravity waves emitted by the convective region are also consistent in amplitude and wavelength with observations. Cloud-top convective activity is also present around the subsolar point. 

In this study we propose to use this convection-resolving to model at noon for two latitudinal cases, the Equator and 75$^{\circ}$, with idealized passive tracers representing \SO and \HO to quantify the impact of the resolved convective layer vertical mixing on the Venusian cloud layer chemistry, and to compare the results with 1D chemical models and observations. The resolved convective plumes give an unprecedented insight into the cloud convective advection and chemistry dynamics.

Our paper is organized as follows. The model is described in Section~\ref{Sec:Model}. In Section~\ref{Sec:Mix}, the impacts of convective motions and gravity waves are presented. The vertical eddy diffusivity is estimated in Section~\ref{Sec:Kzz}. Results are discussed in Section~\ref{Sec:Disc}. Our conclusions are summarized in Section~\ref{Sec:Conc}.

\section{Modelling}
\label{Sec:Model}

\subsection{Mesoscale modelling for Venus}
\label{Sec:Model1}

Our mesoscale model for Venus is based on the dynamical core of the Advanced Research Weather-Weather Research and Forecast (hereinafter referred to as WRF) terrestrial model \citep{Skam08}. The WRF dynamical core integrates the fully compressible non-hydrostatic Navier-Stokes equations on a defined area of the planet. The conservation of the mass, momentum, and entropy are ensured by an explicitly conservative flux-form formulation of the fundamental equations \citep{Skam08}, based on mass-coupled meteorological variables (winds and potential temperature).
The LES mode is used in this study, and the parametrization of the unresolved small-scale eddies is performed using a subgrid-scale prognostic Turbulent Kinetic Energy closure by \cite{Dear72}. This method has been used for Earth convection study \citep{Moen07}, for the Martian atmosphere \citep{Spig10}, and for terrestrial exoplanets \citep{Lefe21}. 

Following the work of \citet{Lefe18} for Venus, the WRF core is coupled to the radiative transfer code of the IPSL Venus General Circulation Model (GCM) \citep{Lebo15} based on \cite{Eyme09} net-exchange rate (NER) formalism computing the energy between the layers prior to the dynamical simulations, separating temperature-independent coefficients from the temperature-dependent Planck functions of the different layers. The cloud model is based on \cite{Haus14} and \cite{Haus15}, with a latitudinal variation of the cloud by setting 5 distinct latitude intervals: 0$^{\circ}$ to 50$^{\circ}$, 50$^{\circ}$ to 60$^{\circ}$, 60$^{\circ}$ to 70$^{\circ}$, 70$^{\circ}$ to 80$^{\circ}$ and 80$^{\circ}$ to 90$^{\circ}$. The cloud model is set prior to the simulations and does not interact with the dynamical features resolved by the model. The solar heating rates are computed using look-up tables \citep{Haus15} of vertical profiles of the solar heating rate as a function of the solar zenith angle. 

\subsection{A Passive tracer approach}
\label{Sec:Model2}

Tracers have been included in the model and are set to represent \SO and \HO. The tracers are not radiatively active. The chemistry, photodissociation, and condensation sources and sinks are modelled by a linear relaxation of the tracer abundance $q$ toward a prescribed vertical profile $q_0 (z)$ with a relaxation timescale $\tau$ computed as :

 \begin{equation}
     \frac{\partial q (x,y,z,t)}{\partial t} = \frac{q_0 (z) - q (x,y,z,t)}{\tau}
 \end{equation}

\begin{figure}
  \centering
  \includegraphics[width=12cm]{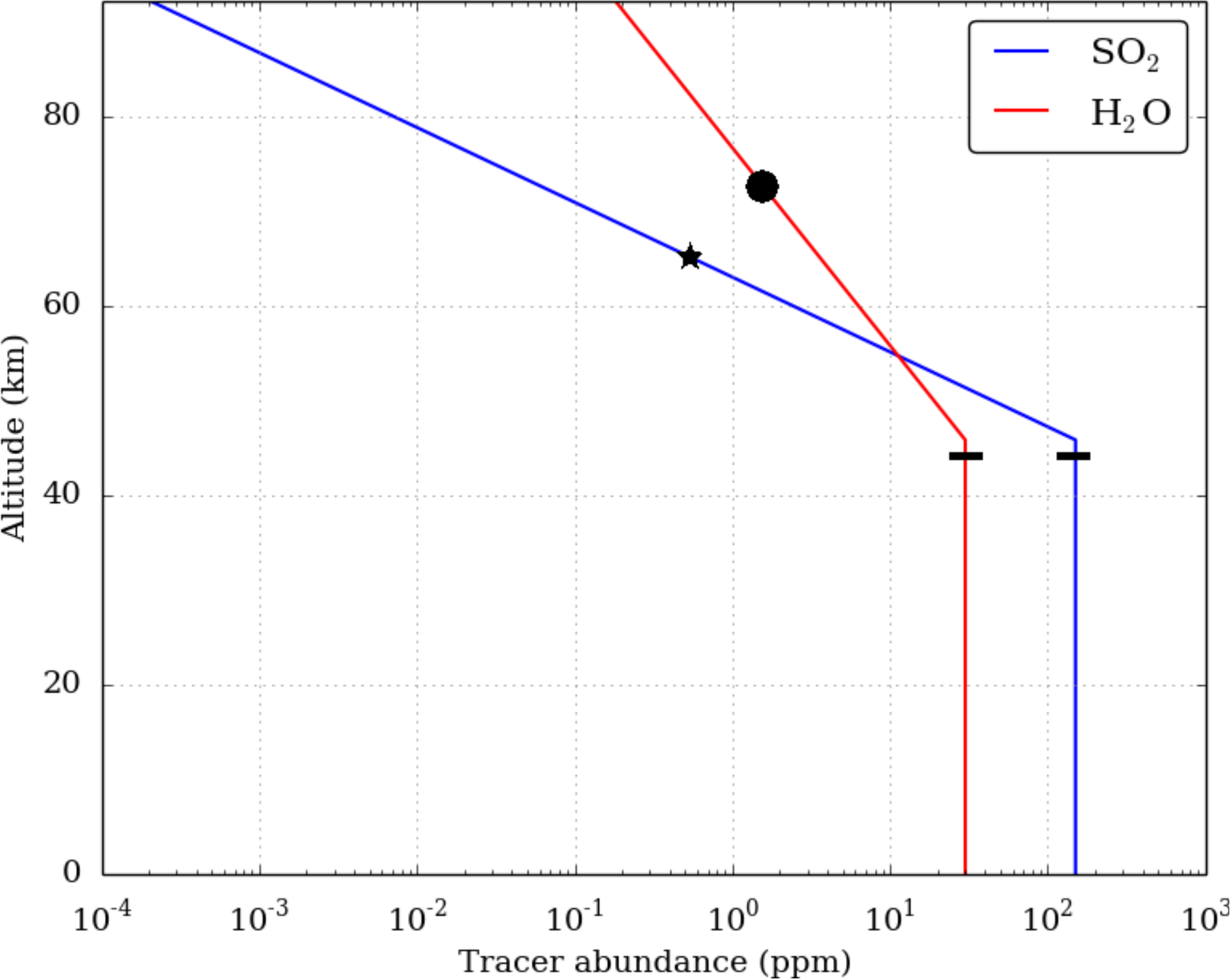}
  \caption{Vertical relaxation tracer abundance profile (ppm). The circle represents the \HO value from \cite{Cott12} and \cite{Fedo16}, the star represents the \SO value from \cite{Encr12,Encr15} and the horizontal line represents tropospheric abundance measurements from \cite{Beza07}.}
  \label{221}
\end{figure}

The two prescribed relaxation profiles are displayed in Fig~\ref{221}. These two profiles are constructed as follows: assuming a constant abundance value below the clouds from spectroscopic measurements between 30 and 40~km \citep{Beza07}, 150~ppm for \SO and 30~ppm for \HO. This value of \SO is consistent with in-situ measurements at the bottom of the cloud deck by \textit{VeGa} 1 and \textit{VeGa} 2 entry probes \citep{Bert96}. From 48~km upwards an exponential decay is assumed going through 0.5~ppm at 65~km, long-term ground observations \citep{Encr12}, for \SO and 2~ppm at 70~km for \HO, VIRTIS and SPICAV measurements \citep{Cott12,Fedo16}. These profiles have been constructed to represent values in the cloud layer, and were constructed for simplicity as constant below the cloud deck and decreasing above. With such simple profiles, there are discrepancies with observations. The value of \SO in the lower cloud is slightly underestimated \citep{Osch21}. Above the clouds, both \SO and \HO are decreasing despite the inversion observed in SPICAV/SOIR \citep{Mahi15} and SPICAV/UV \citep{Evdo21} for \SO, and the very low vertical gradient between 80 and 120~km for \HO observed by SOIR/VEx \citep{Cham20}. The profile of both \SO and \HO outside the cloud layers does not impact the results obtained within the clouds. No source or sinks at the surface or the top boundary are considered. The lateral boundary conditions are doubly periodic. Two latitudinal cases are considered in this study, the Equator and 75$^{\circ}$, to explore the latitudinal variability in terms of convective activity. The relaxation profiles are the same for the two latitudinal cases. Only one local time is considered in this study because of the small diurnal cycle of the convective layer in the model. At night, there is no cloud-top convective layer and this region is similar in terms of dynamics to the high latitude at noon. The chemical timescale of \SO and \HO is not well-constrain over the column, latitudes and local time. Therefore, the relaxation timescale $\tau$ is set to a constant value over the column, with sensitivity tests ranging over values of $10^2$, $10^3$, $10^4$, $10^5$ and $10^6$~s. This broad range of relaxation timescale was chosen to be compared to the dynamical timescale of the convective layer around $10^4$~s (see Section~\ref{Sec:Model3}) with two orders of magnitude above and below this value. \cite{Shao20} reports chemical timescale below $10^4$~s in the upper cloud and as high as $10^8$~s. \cite{Zhan12} reports that the nucleation timescale in the clouds can be below $10^2$~s.  There are no bottom or top boundary conditions for the tracers, the initial profiles for \SO and \HO are considered at equilibrium in regard to chemistry, photochemistry and condensation/vaporization.
 
\subsection{Simulation settings}
\label{Sec:Model3}

\begin{figure}[h!]
  \centering
  \includegraphics[width=16cm]{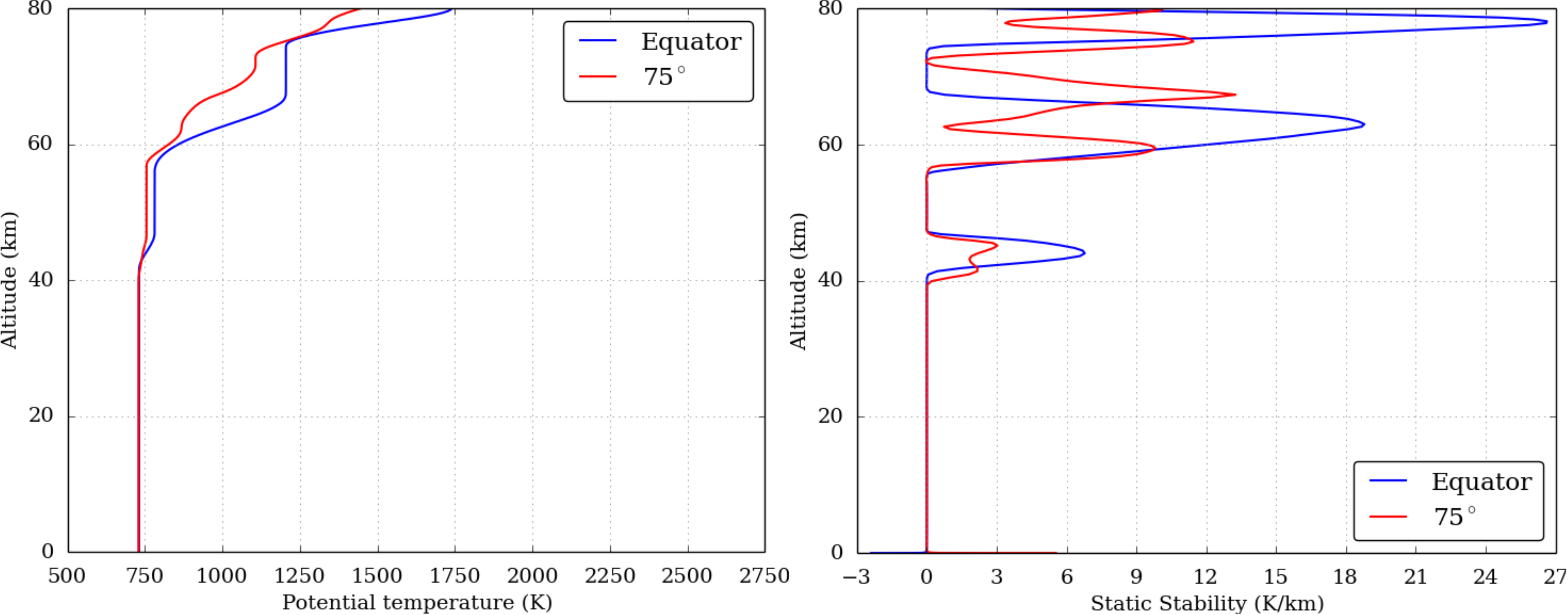}
  \caption{Domain averaged potential temperature (left) and static stability (right) vertical profile from \cite{Lefe18}'s equilibrium state simulations at the Equator and 75$^{\circ}$ at noon. }
  \label{231}
\end{figure}

The simulation settings in terms of resolution and time-step are identical to \cite{Lefe18}'s simulations, with a horizontal resolution of 400~m over 64~km, a vertical domain with 300 points from the surface to 90~km, and a time-step of 1~s. The lateral boundary conditions are doubly periodic. The \cite{Lefe18}'s equilibrium state at the Equator and 75$^{\circ}$ at noon are used as the initial state is shown in Fig~\ref{231}, with the presence of a deep convective layer, between 47 and 56~km at the Equator and between 46.5 and 57~km at high latitude, as well as a hypothetical cloud top convective layer between 66 and 74~km at the Equator. The amplitude of the vertical wind and the diameter of the convective cells are consistent with in-situ measurements of the VeGa balloons \citep{Link86a,Sagd86}. Regarding the gravity waves, with or without the presence of a cloud-top activity, the amplitude of the gravity wave in temperature perturbations and vertical wavelengths are consistent with the radio-occultation measurements \citep{Tell12}, and the horizontal  wavelength is consistent with cloud-top UV observations \citep{Picc14}. With a realistic radiative transfer and incoming solar heating, characteristic of an average UV absorber abundance \citep{Lee19}, the model exhibits a cloud-top convective layer with convective cells diameters consistent with the VMC observations \citep{Mark07,Tito12} that is still speculative but the main hypothesis due to the absence of measurements of corresponding static stability vertical distribution. This cloud-top convective layer has only a limited impact on the deep cloud convective layer because the source, IR heating at cloud base from the troposphere, is unaffected. The thin neutral at 75° of latitude at 7~km is an artefact from the radiative transfer and large-scale heating extracted from the LMD Venus GCM with a much more coarse resolution than the convection-resolving model. To avoid spurious reflection of upward propagating gravity waves on the top boundary of the model, a Rayleigh damping layer is applied over the 8 top kilometres with a damping coefficient of 0.08~s$^{-1}$. The tracers are initialized with relaxation profiles (Fig~\ref{221}) and then advected by the dynamics: the convection, wind shear, and gravity waves. The outputs of the simulations are shown after about two Earth days of simulation.

\section{Mixing}
\label{Sec:Mix}

\begin{figure}[h!]
  \centering
  \includegraphics[width=17cm]{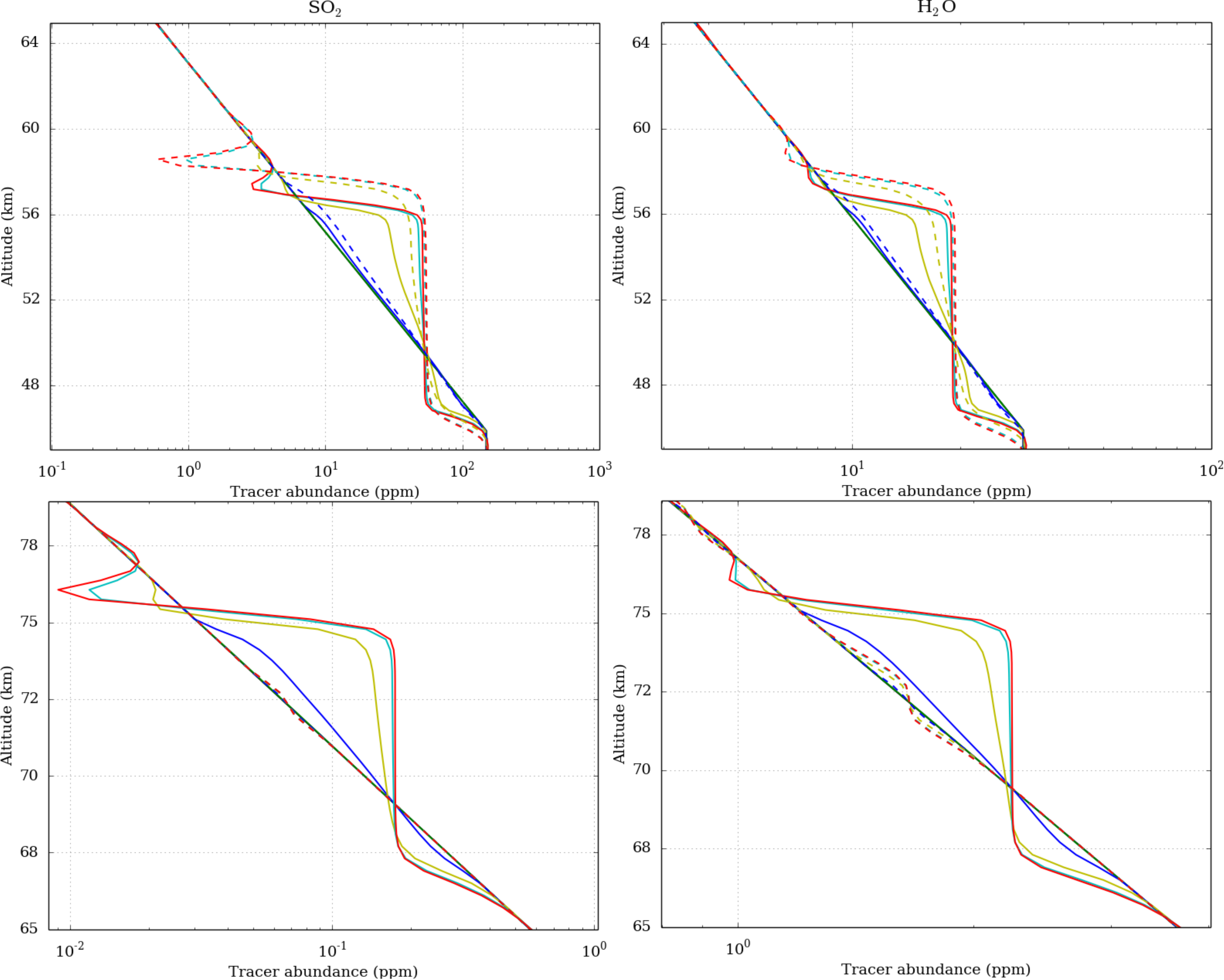}
    \caption{Domain averaged tracer abundance vertical profile of \SO (left column) and \HO (right column) at the initial time (black line), for the deep convective layer (top) and cloud-top convective layer (bottom) for relaxation timescale of 10$^2$~s (green),  10$^3$~s (blue), 10$^4$~s (yellow), 10$^5$~s (cyan) and 10$^6$~s (red) for the Equator (solid line) and 75$^{\circ}$ (dashed line).}
  \label{311}
\end{figure}

Fig~\ref{311} shows the domain averaged vertical profiles of \SO at the Equator and 75$^{\circ}$. The presence of the convective layer, well-mixed, in the deep cloud region is noticeable with a constant tracer abundance value for larger relaxation timescales of $10^5$ and $10^6$~s (Fig~\ref{311}-b). To illustrate the vertical mixing in the cloud layer, a dynamical timescale is defined as follows $\tau_{dyn}$ = H/$\sigma_w$, with H being the scale height and $\sigma_w$ the spatial standard deviation of the vertical wind at a given altitude, and $\tau_{dyn}$ represents the mixing timescale. At the Equator, $\tau_{dyn}$ is equal to about 17000~s at 50~km and equal 6000~s at 70~km. At high latitude $\tau_{dyn}$ is equal to 8000~s at 50~km.

For a relaxation timescale of $10^4$~s, the value of the tracer is not constant in the convective layer but has a significantly steeper vertical gradient compared to the initial state. There is hardly any discernible difference between the initial profile and the tracer abundance profiles for relaxation timescale of $10^2$ and $10^3$~s. When the relaxation timescale is significantly larger than the dynamical timescale $\tau_{dyn}$, like for the $10^5$ and $10^6$~s cases at the Equator, the changes in abundance due to the convection acting faster than the chemical relaxation, and therefore the tracer will be well-mixed. The term chemical relaxation encompasses actual chemical reactions, but also condensation and sublimation for \HO.

On the contrary, when the relaxation timescale is smaller by several orders of magnitude than the dynamical timescale $\tau_{dyn}$, in the $10^2$ and $10^3$~s cases for example, the chemistry will act faster than the perturbations in abundance due to the convection and the abundance will tend to the prescribed value. When the dynamical timescale $\tau_{dyn}$ and the relaxation timescale are close, like in the $10^4$~s case, the tracer abundance will be in-between the two extreme cases. The same competition between the chemical/physical equilibrium and the well-mixed case is also visible at the cloud top. At the Equator, the cases with a relaxation timescale greater than the 6000~s dynamical timescale $\tau_{dyn}$ are well-mixed.

At $75^{\circ}$ of latitude the deep convective layer is slightly thicker with stronger vertical wind, therefore the dynamical timescale $\tau_{dyn}$ is smaller than at the Equator, thus there is a slight difference for the in-between $10^4$~s case, the tracer being closer to the well-mixed case at high latitude. At the cloud top, there is no convective activity, the dynamical timescale becomes very large, and the tracer is always close to the prescribed profile.

Other noticeable features are the abundance drop at the top of the two convective layers and the smaller increase at the bottom of the deep convective layer. These two features are due to the convective entrainment layers, dominated by the updrafts above the convective layers and by the downdrafts below (See Fig~6 of \cite{Lefe17}). The amplitude of these features increases with increasing relaxation timescale; when the convection dominates the vertical mixing, the convective overshoots will have a greater impact.

\subsection{Spatial variability}

\begin{figure}
  \centering
  \includegraphics[width=17cm]{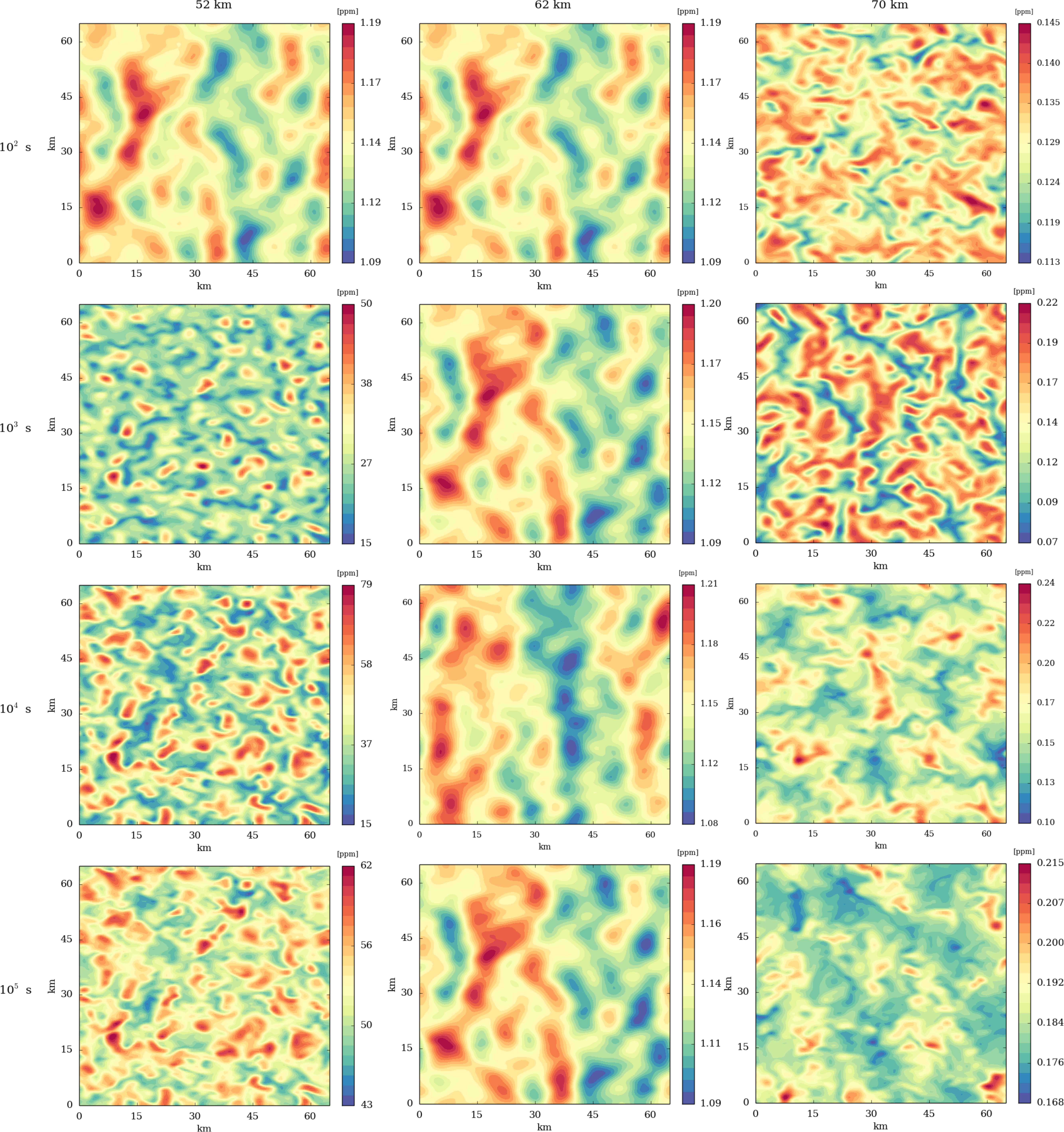}
    \caption{Instantaneous horizontal cross-section of the tracer abundance (ppm) of \SO at the Equator at noon for relaxation timescale values from top to bottom respectively of 10$^2$~s, 10$^3$~s, 10$^4$~s, 10$^5$~s for 52~km (left column), 62~km (middle column) and 70~km (right column).}
  \label{332}
\end{figure}

Fig~\ref{332} displays snapshots of \SO abundance at the same three altitudes in the cloud and for four cases of relaxation timescale at the Equator. The 10$^6$~s case is not shown here for clarity, since it yields very similar results to the 10$^5$~s case. The spatial variability is smaller for extreme relaxation timescale, 10$^2$ and 10$^5$, resulting in smaller intervals ranging less than 10\% in relative value. At 52~km (left column), the effect of the upward convective plume is visible by the high abundance values forming structures up to 10~km in diameter and narrow in-between. At 62~km (middle column), the differences between the different relaxation timescales are hardly noticeable. However, the gravity wave horizontal structure is well-perceptible. As for the cloud-top convective layer (right column), the relaxation timescale will impact in the same way the interval values and the horizontal structure. To precisely measure the horizontal structure of chemical species in the clouds, a resolution of the order of the kilometre is necessary. These trends are similar for \HO (see Fig~\ref{a21} in Appendix~\ref{App2}), and at 75$^{\circ}$ as well, with the presence of gravity waves instead of a convective layer at 70~km for \SO (Fig~\ref{a11} in Appendix~\ref{App1}).

\begin{figure}
  \centering
  \includegraphics[width=16cm]{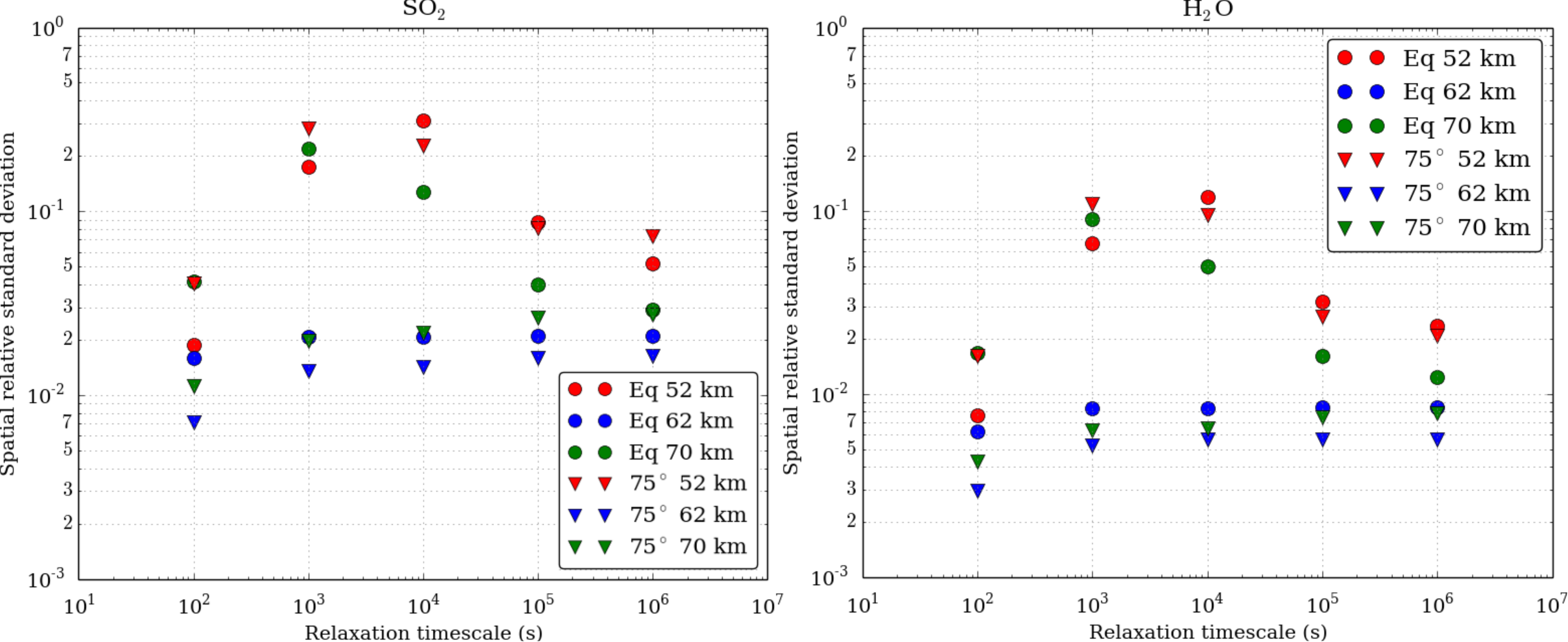}
    \caption{Spatial relative standard deviation for the Equator (circle) and 75$^{\circ}$ (square) at 52~km (red), 62~km (blue) and 70~km (green) for \SO (left) and \HO (right).}
  \label{331}
\end{figure}

Fig~\ref{331} shows the relative standard deviation, the standard deviation divided by the average value at the corresponding altitude, in the horizontal directions for the two latitudinal cases at three altitudes, 52, 62 and 70~km representing respectively the deep convective layer, the gravity waves regions and the cloud-top altitudes, and for both species \SO (left) and \HO (right). The competition between the chemical/physical equilibrium and the convective mixing is readily visible in this figure. At 52~km, As shown in Fig~\ref{332}, for the extreme relaxation timescale cases, 10$^2$, 10$^5$ and 10$^6$~s, the predominance of either the chemistry or the convective mixing over the abundance leads to a spatial relative standard deviation below 0.1, lower by a factor 10 than the in-between cases. Any spatial perturbation would be either mixed by the convection or converted by the chemistry, through a sink or a source. At 62~km, the vertical mixing ensured by the gravity waves is comparatively small (see Fig~\ref{231}), and therefore the spatial abundance disparities are smaller. At 70~km, when there are gravity waves the spatial relative standard deviation is small and when there is the cloud-top convective layer, the spatial relative standard deviation will be maximum for the in-between cases like at 52~km. 
The relaxation timescale also has an impact on the vertical gradient of the tracers inside the convective layer. For the extreme relaxation timescale cases, 10$^2$, 10$^5$ and 10$^6$~s, the vertical gradient will either be strong when the chemistry dominates, or be almost zero when the convective mixing dominates. For intermediate relaxation timescale, there can be the presence of non-monotinic vertical gradient inside the convective layer.

\subsection{Temporal variability}

\begin{figure}
  \centering
  \includegraphics[width=16cm]{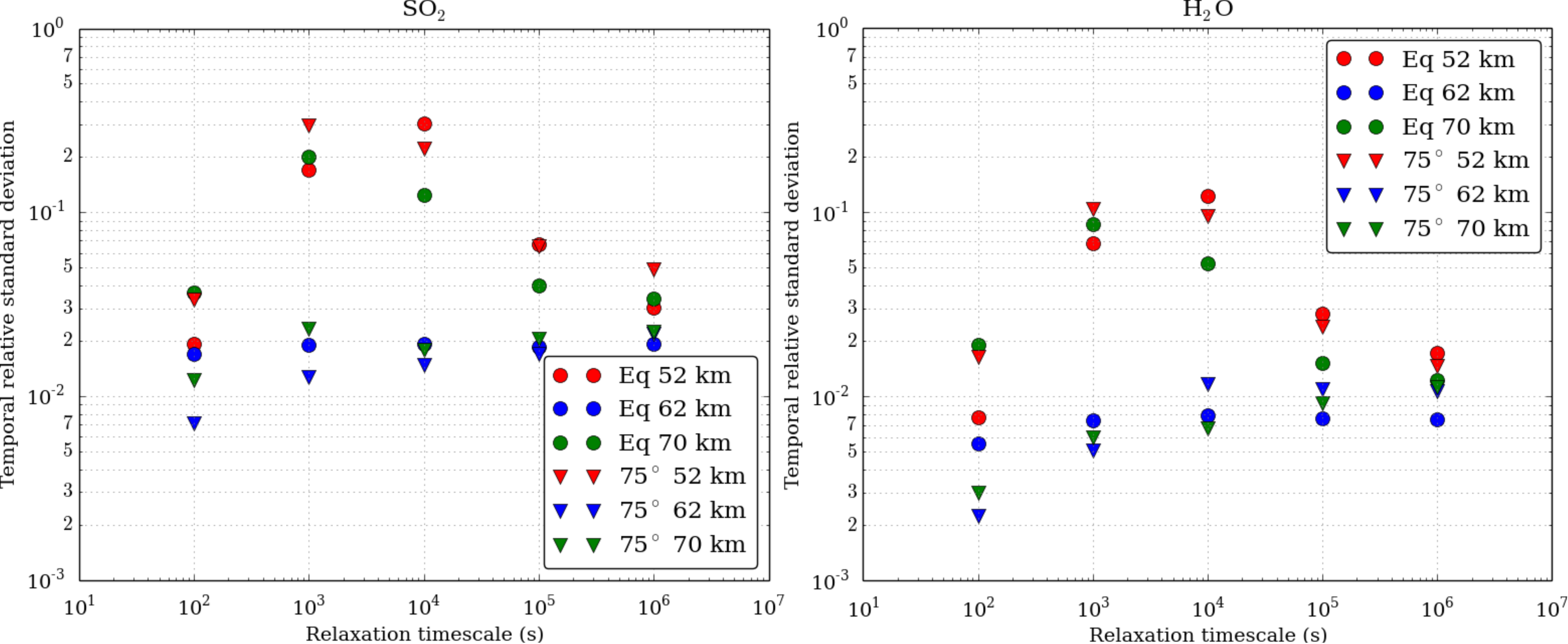}
    \caption{Temporal relative standard deviation for the Equator (circle) and 75$^{\circ}$ (square) at 52~km (red), 62~km (blue) and 70~km (green) for \SO (left) and \HO (right).}
  \label{411}
\end{figure}

Fig~\ref{411} shows the temporal relative standard deviation for the latitudinal cases for the same three altitudes, 52, 62 and 70~km for \SO (left) and \HO (right). The time step is determined as $\min(\tau_{dyn},\tau)/10$, and the standard deviation is computed over a hundred time steps. The competition between the chemical/physical equilibrium and the convective mixing is visible and is of the same order as that of the spatial variability (Fig~\ref{331}). The temporal variability is higher, superior to $0.1$, for the $10^3$ and $10^4$~s cases in the convective layers, at least one order of magnitude over the other location. The physical interpretation is the same as for the spatial variability, related to the competition between $\tau_{dyn}$ and $\tau$.

\section{Vertical eddy diffusion}
\label{Sec:Kzz}

The vertical eddy diffusivity, or $K_{zz}$, is representing the vertical mixing by turbulent processes such as convection or gravity waves, and is used in atmospheric modelling as a parametrization when the turbulence is not resolved, like in 3D GCMs or 1D modelling. For LES models, like the one presented here, with a small resolution, the larger eddies are resolved and no vertical eddy diffusivity is therefore needed.

$K_{zz}$ has been estimated in the Venus atmosphere from observations and modelling, but is still not well-constrained, ranging over several orders of magnitude. The \textit{Pioneer Venus} radio scintillation measurements estimated $K_{zz}$ between $2\cdot 10^{-1}$~m$^2$~s$^{-1}$ at 45~km \citep{Woo82} and 4~m$^2$~s$^{-1}$ at 60~km \citep{Woo81}. Using the values of the vertical wind measured in the deep cloud layer by the \textit{VeGa} balloons, of the order of one meter per second \citep{Sagd86}, $K_{zz}$ was evaluated at $10^3$~m$^2$~s$^{-1}$ at 54~km \citep{Blam86} and between 1 and $10^2$~m$^2$~s$^{-1}$ in the 50-56 km range \citep{Imam01,Gao14}. \cite{Mcgo07} estimated $K_{zz}$ between 1 and $10^3$~m$^2$~s$^{-1}$ in the 50-57 km range. Above the clouds, $K_{zz}$ was estimated using the SOIR/Venus Express CO$_2$ and CO profiles at around 360~m$^2$~s$^{-1}$ from 80~km \citep{Mahi21}.

As the vertical plumes are resolved in LES simulations such as ours, $K_{zz}$ can be estimated from tracers' variations as follows: 
 \begin{equation}
     K_{zz} = -\frac{\langle q'w'\rangle}{\partial \langle q\rangle/\partial z}
     \label{eq22}
 \end{equation}
 
with $q$ a tracer mixing ratio, primed quantities representing perturbations relative to the domain averaged values, and bracket quantities representing domain averaged values. 
 
\begin{figure}[H]
  \centering
  \includegraphics[width=14cm]{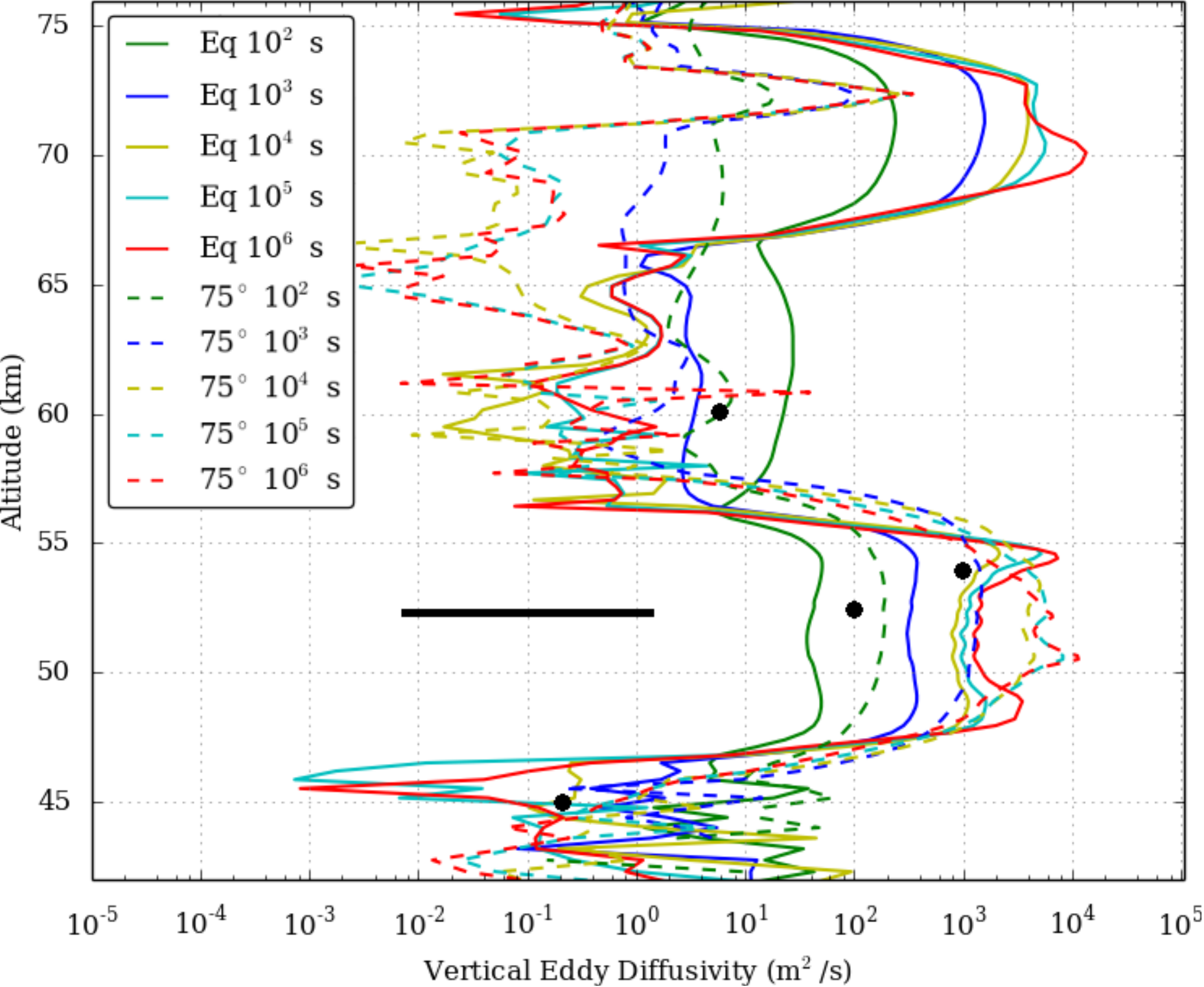}
    \caption{Vertical profiles of the vertical eddy diffusivity (m$^2$~s$^{-1}$) in the Venus cloud region calculated with the \SO tracer for relaxation timescale of 10$^2$~s (green), 10$^3$~s (blue), 10$^4$~s (yellow), 10$^5$~s (cyan) and 10$^6$~s (red) for the Equator (solid line) and 75$^{\circ}$ (dashed line). The black circles represent the previous estimations (see Table~\ref{T1}) and the black line represents the range of values used for the cloud convective layer in recent chemical models \cite{Kras12,Bier20,Rimm21}.}
  \label{51}
\end{figure}

Fig~\ref{51} displays the vertical eddy diffusivity profiles for the two latitudinal cases and the four relaxation timescale considered with \SO. Below the clouds, the value of $K_{zz}$ is in the order of $10^{-1}$~m$^2$~s$^{-1}$ consistent with \citep{Woo82}. In the deep convective layer, $K_{zz}$ increases and reaches values up to $10^{3}$~m$^2$~s$^{-1}$ at the Equator and $10^{4}$~m$^2$~s$^{-1}$ at high latitudes. The order of magnitude is consistent with estimation using Prandtl mixing-length theory \citep{Lind71}. The values for relaxation timescale equal to $10^{4}$, $10^{5}$ and $10^{6}$~s at the Equator are consistent with in-situ measurements from \textit{VeGa} balloons at the same latitudes \citep{Blam86}. Most of the values between 50 and 57~km are consistent with the estimation of \cite{Imam01,Gao14} and \cite{Mcgo07}. However, at high latitudes the vertical eddy diffusivity values exceed these estimations, reaching $10^{4}$~m$^2$~s$^{-1}$, due to a thicker convective layer, consistent with VeRa measurements \citep{Tell09}. The relaxation has an important impact on the vertical mixing, and changes the value of the vertical eddy diffusivity by several orders of magnitude, for high $\tau$ values the tracer is well-mixed with a small vertical gradient in the convective layer, resulting in a high vertical eddy diffusivity. Above, $K_{zz}$ decreases to $1$~m$^2$~s$^{-1}$, consistent with in-situ measurements \citep{Woo81} at 60~km. At noon near the Equator, the gravity waves resolved by the model are trapped between the two convective layers and have a stronger amplitude. The vertical eddy diffusivity is therefore stronger than at higher latitudes, with an opposite effect upon the relaxation timescale, where $K_{zz}$ is stronger for a lower value of $\tau$. At the cloud top at noon, the vertical eddy diffusivity strongly increases to values up to $10^4$~m$^2$~s$^{-1}$ due to the cloud-top convective layer generated by solar heating absorption from the unknown UV absorber \citep{Lefe18}, where the relaxation has an important effect similarly as the deep cloud convective layer. The presence of a cloud-top convective layer is still speculative, however the turbulence resolved by the model provides an insight of the impact of such convective activity and trapped gravity waves. Without the presence of this cloud-top convective layer, the mixing of the gravity waves at the Equator above the deep convection would be equivalent to the mixing at 75$^{\circ}$ of latitude in the gravity waves region. Regarding the local time variability, the deep convective layer is deeper at night \cite{Imam17,Lefe18}. Therefore, the mixing would be stronger. The small increase for 75$^{\circ}$ of latitude at 72~km is an artefact mentioned in Section~\ref{Sec:Model3}. The vertical eddy diffusivity profiles were also estimated using \HO, and the orders of magnitude are consistent between the two species. The estimation of the vertical eddy diffusion for heat, using potential temperature instead of tracer abundance in Equation~\ref{eq22}, gives a value between $10^{3}$ and $10^{4}$~m$^2$~s$^{-1}$ for the Equator in the deep cloud convective region and around $10^{4}$~m$^2$~s$^{-1}$ for the cloud-top convective layer, and between $10^{4}$ and $10^{5}$~m$^2$~s$^{-1}$ for high latitudes. 
Above the clouds and below the Rayleigh damping layer, $K_{zz}$ is estimated from 1 to 10~m$^2$~s$^{-1}$, at least one order of magnitude lower than the values estimated from 80~km \citep{Mahi21}. The small-scale turbulence only plays a minor role in the mixing in this region. The present estimation of the vertical eddy diffusivity coefficient and comparison with previous estimations are summarized in Table~\ref{T1}.

\begin{table}
\center
\begin{tabular}{|l|c|c|}
\hline
$K_{zz}$ (m$^2$~s$^{-1}$) & Previous estimations & This study \\
\hline
at 45~km & $2\cdot 10^{-1}$ \citep{Woo82}& $10^{-1}$-1 \\
\hline
at 54~km & $10^3$ \citep{Blam86} & 50-4000\\
50-56~km & 1-$10^2$ \citep{Imam01} & 20-10 000\\
50-57~km & 1-$10^3$ \citep{Mcgo07} & $10^{-1}$-10 000\\
\hline
at 60~km & 4 \citep{Woo81} & $10^{-1}$-20\\
\hline
\end{tabular}
\caption{Vertical eddy diffusivity K$_{zz}$ (m$^2$~s$^{-1}$) estimations.}
\label{T1}
\end{table}

\section{Discussion}
\label{Sec:Disc}

The \cite{Kras12} 1D model uses an eddy diffusion of 1~m$^2$~s$^{-1}$ in the convective layer, at least 10 times lower than the present study, constant up to a specific altitude set to 55, 60 or 65~km, and then increases linearly. Both modelled \SO and \HO abundance are sensitive to small variations of eddy diffusion, not constant in the convective layer, suggesting a short chemical timescale, inferior to 10$^4$~s. The 55 and 60~km top altitude of the deep eddy diffusion constant value results in a better agreement with mesospheric measurement values. This eddies diffusion constant value depth is consistent with the convective layer depth in this study.

The 1D model of \cite{Bier20} defined several eddy diffusion scenarios for the convective layer mixing parametrization, from 0.1 to 2~m$^2$~s$^{-1}$ nominal eddy diffusion value, at least one order of magnitude lower than the present study. With the nominal eddy diffusion set-up, constant from the surface to 60~km, the transport timescale ranges from 10 years at the surface to a few months at 90~km, orders of magnitude larger in the convective layer than the convective dynamical timescale $\tau_{dyn}$. However, \SO exhibits in their model a very well-mixed profile below 60~km, a step decrease and then another region with a very low vertical gradient up to 110~km, inconsistent with the observations from \textit{Venus Express} \citep{Mahi15,Vand17a,Evdo21}. The exploration of the eddy diffusion values and profile in the convective layer leads to a better agreement with the observations of \SO abundance in the mesosphere with a lower value. \cite{Rimm21} also tested several vertical eddy diffusion values in the convective layer, between, used also 0.01 to 1~m$^2$~s$^{-1}$. However, these tested eddy diffusion values are inconsistent with estimations from in-situ wind and cloud microphysics, and the values presented in Section~\ref{Sec:Kzz}.

The chemical models of \cite{Yung09} and \cite{Zhan12} and \cite{Shao20} simulate the atmosphere from 58~km, therefore above the convective layer. \cite{Shao20} showed that the temporal variations of \SO and \HO are linked to the lower atmospheric processes, such as convection. No chemistry model considers the cloud-top convection hypothesis through the inclusion of variable $K_{zz}$ profiles. In the IPSL Venus GCM, $K_{zz}$ reaches values above 10~m$^2$~s$^{-1}$ between 48 and 57~km \citep{Lebo17}, the lowest value estimated with the present convection resolving model.

The minor gaseous species have been measured mainly in the mesosphere by Venus Express above 60~km, e.g. \citet{Marc20} for \SO and \citet{Cott12,Fedo16,Vand17a,Cham20} for \HO, and around 60~km by ground-based telescopes \citep{Encr19,Encr20} and Hubble Space Telescope \citep{Jess15,Jess20}. The spatial resolution of \textit{Venus Express} VIRTIS and SPICAV instruments can be around a few tens of kilometre at best \citep{Dros07,Bert07b}, while the spatial resolution is about 100~km for the ground-based observations and between 20 and 60~km for the Hubble measurements. The uncertainty about the height of the probed regions for both IR and UV observations makes the comparison between the present model and observations delicate.

A short-time variability of two hours for \SO was measured from the ground \citep{Encr16} at around 60~km, consistent with the dynamical timescale of the convective layer presented here. The probability of this variability is stronger at night \citep{Encr19}, consistent with the local variability of the convective layer in the model \citep{Lefe18} and observations \citep{Imam17}. \textit{VeGa} 1 and \textit{VeGa} 2 entry probes measured \SO between 60~km and the ground \citep{Bert96}. Inside the convective layer, a strong non-monotonic vertical gradient was measured, consistent with the vertical gradient for intermediate relaxation timescale of \SO in the same region but with a higher amplitude.

At cloud-top altitudes, observed \SO does not show clear signs of a vertically well-mixed layer with a large set of values of the vertical gradient \citep{Vand17a}, suggesting a short chemical timescale (inferior to 10$^3$~s) or no cloud-top convection layer. However, \SO was measured at the top of the cloud layer by SPICAV-UV, a higher variability is observed below 30$^{\circ}$ of latitude with an order of magnitude consistent with the presence of a cloud-top convective layer \citep{Marc20}, in the same latitude range as the cellular features observed by VMC \citep{Tito12}, consistent with a relaxation timescale between 10$^3$ and 10$^4$~s. At these latitudes, the abundance of observed \SO is smaller around noon correlated with an increase of the unknown UV absorber, whose solar heating is needed to generate the cloud-top convective layer. This minimum of \SO at noon was also observed from the ground \citep{Encr19,Encr20}.

In the same altitude range, \HO was measured by VEX/VIRTIS in the turbulent-like cloud features \citep{Cott12}. The horizontal abundance variation in this area is about 1 ppm, consistent with our results (Fig~\ref{331} and Fig~\ref{a21}) with a relaxation timescale of 10$^3$ and 10$^4$~s.

SPICAV measured \HO in the upper cloud \citep{Fedo16}, between 61 and 59~km (the lowest altitude probed) the vertical gradient is very weak, consistent with a well-mixed layer. A small vertical gradient of \HO is needed for better retrievals of HDO ground observations \citep{Encr13}, whereas a significant vertical gradient is used for \SO. \citep{Encr16} measured a higher variability for \SO than for \HO, suggesting again a well-mixed layer for water. The variations observed at the cloud top show no signs of a diurnal cycle at low latitudes, suggesting also a short chemical timescale (inferior to 10$^3$~s) or no convection layer.

\section{Conclusion}
\label{Sec:Conc}
This study is the first one to investigate the vertical mixing in the Venusian cloud layer on the chemistry using 3D resolved convective plumes. Using an idealized passive tracer to represent \SO and \HO, vertical mixing of those species by the convection was estimated depending on the chemical timescale. A small chemical timescale compared to the convective timescale, about $10^3$~s or lower, will limit the vertical turbulent mixing. On the contrary, a chemical timescale value superior to the convective timescale, about $10^5$~s or higher, will lead to a well-mixed layer, i.e. a small vertical gradient in the convective region. The convective region is organized as updraft polygonal cells, and the spatial and temporal variability through the cloud is estimated. These variabilities are maximal in the convective regions for a convective chemical timescale equivalent to the convective timescale. The horizontal structure of the \SO and \HO is composed of small-scale structures due to the convective cells or gravity waves as small as a few kilometres in the convective regions. In the gravity wave area, the horizontal structure of the tracer is governed by the gravity wave horizontal wavelength. The mixing on tracers of the hypothetical cloud-top convective layer has been estimated, with a smaller convective timescale, the convective plumes can engender small-scale structures. With the presence of such a cloud-top convective layer, the gravity waves have larger amplitudes and exhibit a stronger mixing. The higher values of the mixing in the upper cloud layer could help to infer the presence of a cloud-top convective layer.

From the resolving plumes, the vertical eddy diffusion has been estimated in the cloud region. For the deep cloud convection, $K_{zz}$ ranges from $10^{1}$ to $10^{4}$~m$^2$~s$^{-1}$, consistent with in-situ and modelling estimation. There is an impact of the relaxation timescale, the stronger it is, the higher $K_{zz}$ is, and an impact of the latitudinal convection depth variability. In the gravity waves region, the vertical eddy diffusion is lower, and increases again at the cloud top at the subsolar point due to the cloud-top convective activity.

Previous 1D chemical models \cite{Yung09,Bier20,Rimm21} found no mixed layers for \SO and \HO. However, the various vertical eddy diffusion profiles and values tested are not consistent with the values of the present study and convective mixing is underestimated. These models show that vertical eddy diffusion is important for the mesospheric abundance of the chemical species, there is a need for new 1D chemical simulations with a more realistic convective mixing.

The limited size of the domain and the simplicity of the representation of the chemistry makes comparison with the observations difficult. Nevertheless, there are interesting points to comment on. A two-hour variability of \SO around 60~km measured from the ground is compatible with the dynamical timescale of the convective layer. The horizontal abundance variation of \HO in the turbulent-looking cloud top is consistent with the spatial variability of the subsolar cloud-top convection. The weak vertical gradient of \HO between 61 and 58~km is consistent with convective layer vertical mixing. The vertical profiles of \SO and \HO do not indicate clear signs of cloud-top convective mixing. However, the cloud-top variability of \SO around the subsolar point is consistent with the presence of a cloud-top convective layer.

\subsection*{Future improvements}

A wider horizontal domain is needed to be able to compare observations with more accuracy. The passive tracer method used in this study is idealized, a proper chemistry and microphysics scheme are needed for a more realistic model, and to be able to quantify the impact of the convective activity on cloud formation and cloud opacity or to study the impact of turbulence on coupled particles trend, like the anti-correlated variations of \HO and \SO predicted by 1D chemical model \citep{Park15,Shao20} and observed from the ground \citep{Encr20}.

The cloud-top convective activity resolved in the model is generated by the unknown UV absorber. However, its abundance distribution is not well-constrained and this uncertainty could have a strong effect on the cloud-top convection depth and vertical mixing. The observed variations of this unknown absorber \citep{Lee19} need to be taken into account for a better understanding of this cloud-top convective layer.

This study focuses on small-scale turbulence, a similar study needs to be conducted for the mesoscale features observed in the clouds with \textit{Akatsuki}, like the vertical propagation mountain \citep{Kouy17,Lefe20} and the other new cloud morphologies \citep{Pera19}.

Convective-resolving studies have been used for Earth \citep{Rio10} and Mars \citep{Cola13} atmospheres to improve the parametrization of the convection in the GCMs, the present study could be the starting point of such methodology for a more sophisticated approach for the mixing of heat, momentum, and species in the Venus cloud layer.

\appendix
\begin{appendices}

\newpage
\section{\SO Maps}
\label{App1}

\begin{figure}[h!]
  \centering
  \includegraphics[width=\textwidth]{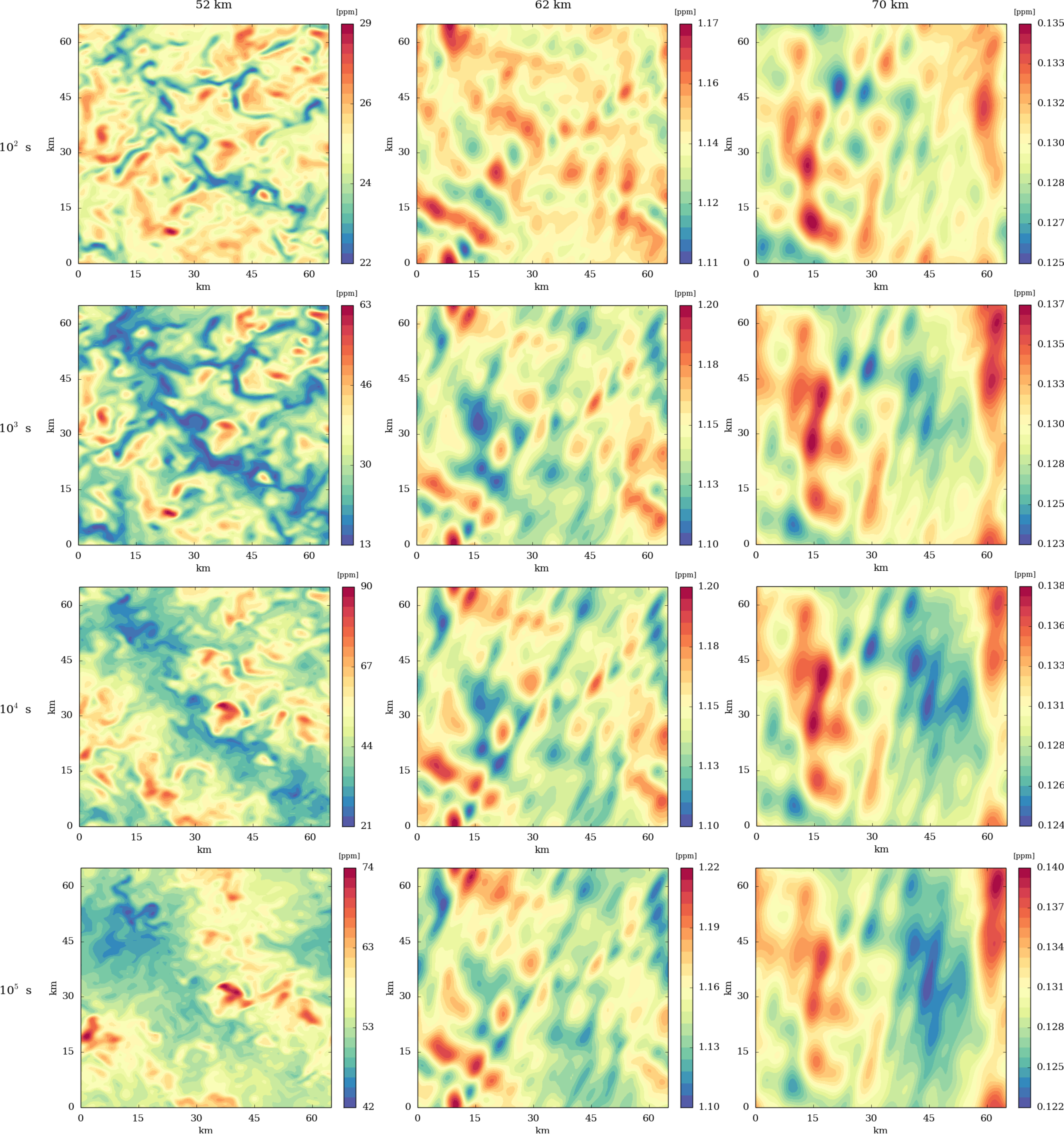}
    \caption{Instantaneous horizontal cross-section of the tracer abundance (ppm) of \SO at 75$^{\circ}$ at noon for relaxation timescale values from top to bottom respectively of 10$^2$~s, 10$^3$~s, 10$^4$~s, 10$^5$~s for 52~km (left column), 62~km (middle column) and 70~km (right column).}
  \label{a11}
\end{figure}

\newpage
\section{\HO Maps}
\label{App2}

\begin{figure}[h!]
  \centering
  \includegraphics[width=\textwidth]{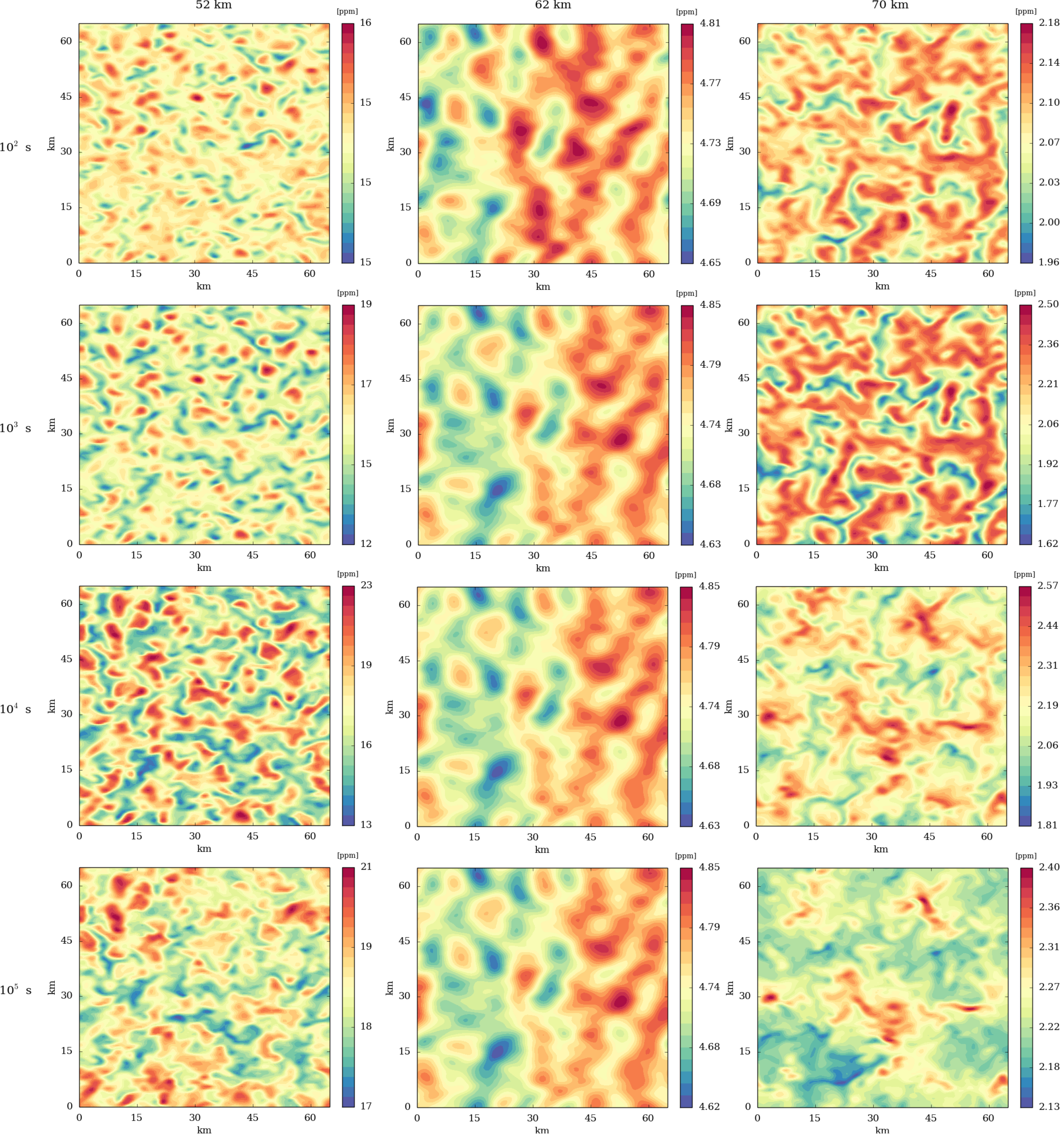}
    \caption{Instantaneous horizontal cross-section of the tracer abundance (ppm) of \HO at the Equator at noon for relaxation timescale values from top to bottom respectively of 10$^2$~s, 10$^3$~s, 10$^4$~s, 10$^5$~s for 52~km (left column), 62~km (middle column) and 70~km (right column).}
  \label{a21}
\end{figure}

\end{appendices}

\newpage
\section*{Acknowledgements}
The authors would like to thank Th\'er\`ese Encrenaz for the discussions on observations, and Aymeric Spiga and S\'ebastien Lebonnois for their help on the manuscript. The authors would like to thank the two anonymous reviewers that helped to improve this study. ML acknowledges funding from the European Research Council (ERC) under the European Union’s Horizon 2020 research and innovation program (grant agreement No. 740963/EXOCONDENSE). ML would like to acknowledge the use of the University of Oxford Advanced Research Computing (ARC) facility in carrying out this work. \url{http://dx.doi.org/10.5281/zenodo.22558}. EM acknowledges support from CNES and ESA for science activities related to the VenSpec-U instrumental development and scientific activities onboard \emph{EnVision} (ESA Cosmic Vision M5 mission). Simulation results used to obtain the figures in this paper are available in the open online repository \url{https://figshare.com/s/39ad796d4281fab08511}. Full simulation results performed in this paper are available upon reasonable request.


\begin{thebibliography}{}

\bibitem[{Ando} et~al., 2020]{Ando20}
{Ando}, H., {Imamura}, T., {Tellmann}, S., {P{\"a}tzold}, M., {H{\"a}usler},
  B., {Sugimoto}, N., {Takagi}, M., {Sagawa}, H., {Limaye}, S., {Matsuda}, Y.,
  {Choudhary}, R.~K., and {Antonita}, M. (2020).
\newblock {Thermal structure of the Venusian atmosphere from the sub-cloud
  region to the mesosphere as observed by radio occultation}.
\newblock {\em Scientific Reports}, 10:3448.

\bibitem[{Ando} et~al., 2018]{Ando18}
{Ando}, H., {Takagi}, M., {Fukuhara}, T., {Imamura}, T., {Sugimoto}, N.,
  {Sagawa}, H., {Noguchi}, K., {Tellmann}, S., {P{\"a}tzold}, M.,
  {H{\"a}usler}, B., {Murata}, Y., {Takeuchi}, H., {Yamazaki}, A., {Toda}, T.,
  {Tomiki}, A., {Choudhary}, R., {Kumar}, K., {Ramkumar}, G., and {Antonita},
  M. (2018).
\newblock {Local Time Dependence of the Thermal Structure in the Venusian
  Equatorial Upper Atmosphere: Comparison of Akatsuki Radio Occultation
  Measurements and GCM Results}.
\newblock {\em Journal of Geophysical Research (Planets)}, 123(9):2270--2280.

\bibitem[{Belton} et~al., 1976]{Belt76b}
{Belton}, M.~J.~S., {Smith}, G.~R., {Schubert}, G., and {del Genio}, A.~D.
  (1976).
\newblock {Cloud patterns, waves and convection in the Venus atmosphere}.
\newblock {\em Journal of Atmospheric Sciences}, 33:1394--1417.

\bibitem[{Bertaux} et~al., 2007]{Bert07b}
{Bertaux}, J.-L., {Nevejans}, D., {Korablev}, O., {Villard}, E.,
  {Qu{\'e}merais}, E., {Neefs}, E., {Montmessin}, F., {Leblanc}, F., {Dubois},
  J.~P., {Dimarellis}, E., {Hauchecorne}, A., {Lef{\`e}vre}, F., {Rannou}, P.,
  {Chaufray}, J.~Y., {Cabane}, M., {Cernogora}, G., {Souchon}, G., {Semelin},
  F., {Reberac}, A., {Van Ransbeek}, E., {Berkenbosch}, S., {Clairquin}, R.,
  {Muller}, C., {Forget}, F., {Hourdin}, F., {Talagrand}, O., {Rodin}, A.,
  {Fedorova}, A., {Stepanov}, A., {Vinogradov}, I., {Kiselev}, A.,
  {Kalinnikov}, Y., {Durry}, G., {Sandel}, B., {Stern}, A., and {G{\'e}rard},
  J.~C. (2007).
\newblock {SPICAV on Venus Express: Three spectrometers to study the global
  structure and composition of the Venus atmosphere}.
\newblock {\em Planetary and Space Science}, 55(12):1673--1700.

\bibitem[{Bertaux} et~al., 1996]{Bert96}
{Bertaux}, J.-L., {Widemann}, T., {Hauchecorne}, A., {Moroz}, V.~I., and
  {Ekonomov}, A.~P. (1996).
\newblock {VEGA 1 and VEGA 2 entry probes: An investigation of local UV
  absorption (220-400 nm) in the atmosphere of Venus (SO$_{2}$, aerosols, cloud
  structure)}.
\newblock {\em Journal of Geophysical Research}, 101(E5):12709--12746.

\bibitem[{B{\'e}zard} and {de Bergh}, 2007]{Beza07}
{B{\'e}zard}, B. and {de Bergh}, C. (2007).
\newblock {Composition of the atmosphere of Venus below the clouds}.
\newblock {\em Journal of Geophysical Research (Planets)}, 112(E4):E04S07.

\bibitem[{Bierson} and {Zhang}, 2020]{Bier20}
{Bierson}, C.~J. and {Zhang}, X. (2020).
\newblock {Chemical Cycling in the Venusian Atmosphere: A Full Photochemical
  Model From the Surface to 110 km}.
\newblock {\em Journal of Geophysical Research (Planets)}, 125(7):e06159.

\bibitem[{Blamont} et~al., 1986]{Blam86}
{Blamont}, J.~E., {Young}, R.~E., {Seiff}, A., {Ragent}, B., {Sagdeev}, R.,
  {Linkin}, V.~M., {Kerzhanovich}, V.~V., {Ingersoll}, A.~P., {Crisp}, D.,
  {Elson}, L.~S., {Preston}, R.~A., {Golitsyn}, G.~S., and {Ivanov}, V.~N.
  (1986).
\newblock {Implications of the VEGA balloon results for Venus atmospheric
  dynamics}.
\newblock {\em Science}, 231:1422--1425.

\bibitem[{Chamberlain} et~al., 2020]{Cham20}
{Chamberlain}, S., {Mahieux}, A., {Robert}, S., {Piccialli}, A., {Trompet}, L.,
  {Vandaele}, A.~C., and {Wilquet}, V. (2020).
\newblock {SOIR/VEx observations of water vapor at the terminator in the Venus
  mesosphere}.
\newblock {\em Icarus}, 346:113819.

\bibitem[{Cola{\"\i}tis} et~al., 2013]{Cola13}
{Cola{\"\i}tis}, A., {Spiga}, A., {Hourdin}, F., {Rio}, C., {Forget}, F., and
  {Millour}, E. (2013).
\newblock {A thermal plume model for the Martian convective boundary layer}.
\newblock {\em Journal of Geophysical Research (Planets)}, 118(7):1468--1487.

\bibitem[{Cottini} et~al., 2012]{Cott12}
{Cottini}, V., {Ignatiev}, N.~I., {Piccioni}, G., {Drossart}, P., {Grassi}, D.,
  and {Markiewicz}, W.~J. (2012).
\newblock {Water vapor near the cloud tops of Venus from Venus Express/VIRTIS
  dayside data}.
\newblock {\em Icarus}, 217:561--569.

\bibitem[{Counselman} et~al., 1980]{Coun80}
{Counselman}, C.~C., {Gourevitch}, S.~A., {King}, R.~W., {Loriot}, G.~B., and
  {Ginsberg}, E.~S. (1980).
\newblock {Zonal and meridional circulation of the lower atmosphere of Venus
  determined by radio interferometry}.
\newblock {\em Journal of Geophysical Research}, 85:8026--8030.

\bibitem[{Deardorff}, 1972]{Dear72}
{Deardorff}, J.~W. (1972).
\newblock {Numerical Investigation of Neutral and Unstable Planetary Boundary
  Layers.}
\newblock {\em Journal of Atmospheric Sciences}, 29:91--115.

\bibitem[{Drossart} et~al., 2007]{Dros07}
{Drossart}, P., {Piccioni}, G., {Adriani}, A., {Angrilli}, F., {Arnold}, G.,
  {Baines}, K.~H., {Bellucci}, G., {Benkhoff}, J., {B{\'e}zard}, B., {Bibring},
  J.~P., {Blanco}, A., {Blecka}, M.~I., {Carlson}, R.~W., {Coradini}, A., {Di
  Lellis}, A., {Encrenaz}, T., {Erard}, S., {Fonti}, S., {Formisano}, V.,
  {Fouchet}, T., {Garcia}, R., {Haus}, R., {Helbert}, J., {Ignatiev}, N.~I.,
  {Irwin}, P.~G.~J., {Langevin}, Y., {Lebonnois}, S., {Lopez-Valverde}, M.~A.,
  {Luz}, D., {Marinangeli}, L., {Orofino}, V., {Rodin}, A.~V., {Roos-Serote},
  M.~C., {Saggin}, B., {Sanchez-Lavega}, A., {Stam}, D.~M., {Taylor}, F.~W.,
  {Titov}, D., {Visconti}, G., {Zambelli}, M., {Hueso}, R., {Tsang}, C.~C.~C.,
  {Wilson}, C.~F., and {Afanasenko}, T.~Z. (2007).
\newblock {Scientific goals for the observation of Venus by VIRTIS on ESA/Venus
  express mission}.
\newblock {\em Planetary and Space Science}, 55(12):1653--1672.

\bibitem[{Encrenaz} et~al., 2019]{Encr19}
{Encrenaz}, T., {Greathouse}, T.~K., {Marcq}, E., {Sagawa}, H., {Widemann}, T.,
  {B{\'e}zard}, B., {Fouchet}, T., {Lef{\`e}vre}, F., {Lebonnois}, S.,
  {Atreya}, S.~K., {Lee}, Y.~J., {Giles}, R., and {Watanabe}, S. (2019).
\newblock {HDO and SO$_{2}$ thermal mapping on Venus. IV. Statistical analysis
  of the SO$_{2}$ plumes}.
\newblock {\em Astronomy \& Astrophysics}, 623:A70.

\bibitem[{Encrenaz} et~al., 2020]{Encr20}
{Encrenaz}, T., {Greathouse}, T.~K., {Marcq}, E., {Sagawa}, H., {Widemann}, T.,
  {B{\'e}zard}, B., {Fouchet}, T., {Lef{\`e}vre}, F., {Lebonnois}, S.,
  {Atreya}, S.~K., {Lee}, Y.~J., {Giles}, R., {Watanabe}, S., {Shao}, W.,
  {Zhang}, X., and {Bierson}, C.~J. (2020).
\newblock {HDO and SO$_{2}$ thermal mapping on Venus. V. Evidence for a
  long-term anti-correlation}.
\newblock {\em Astronomy \& Astrophysics}, 639:A69.

\bibitem[{Encrenaz} et~al., 2016]{Encr16}
{Encrenaz}, T., {Greathouse}, T.~K., {Richter}, M.~J., {DeWitt}, C.,
  {Widemann}, T., {B{\'e}zard}, B., {Fouchet}, T., {Atreya}, S.~K., and
  {Sagawa}, H. (2016).
\newblock {HDO and SO$_{2}$ thermal mapping on Venus. III. Short-term and
  long-term variations between 2012 and 2016}.
\newblock {\em Astronomy \& Astrophysics}, 595:A74.

\bibitem[{Encrenaz} et~al., 2013]{Encr13}
{Encrenaz}, T., {Greathouse}, T.~K., {Richter}, M.~J., {Lacy}, J., {Widemann},
  T., {B{\'e}zard}, B., {Fouchet}, T., {deWitt}, C., and {Atreya}, S.~K.
  (2013).
\newblock {HDO and SO$_{2}$ thermal mapping on Venus. II. The SO$_{2}$ spatial
  distribution above and within the clouds}.
\newblock {\em Astronomy \& Astrophysics}, 559:A65.

\bibitem[{Encrenaz} et~al., 2012]{Encr12}
{Encrenaz}, T., {Greathouse}, T.~K., {Roe}, H., {Richter}, M., {Lacy}, J.,
  {B{\'e}zard}, B., {Fouchet}, T., and {Widemann}, T. (2012).
\newblock {HDO and SO$_{2}$ thermal mapping on Venus: evidence for strong
  SO$_{2}$ variability}.
\newblock {\em Astronomy \& Astrophysics}, 543:A153.

\bibitem[{Encrenaz} et~al., 2015]{Encr15}
{Encrenaz}, T., {Moreno}, R., {Moullet}, A., {Lellouch}, E., and {Fouchet}, T.
  (2015).
\newblock {Submillimeter mapping of mesospheric minor species on Venus with
  ALMA}.
\newblock {\em Planetary and Space Science}, 113:275--291.

\bibitem[{Evdokimova} et~al., 2021]{Evdo21}
{Evdokimova}, D., {Belyaev}, D., {Montmessin}, F., {Korablev}, O., {Bertaux},
  J.~L., {Verdier}, L., {Lef{\`e}vre}, F., and {Marcq}, E. (2021).
\newblock {The Spatial and Temporal Distribution of Nighttime Ozone and Sulfur
  Dioxide in the Venus Mesosphere as Deduced From SPICAV UV Stellar
  Occultations}.
\newblock {\em Journal of Geophysical Research (Planets)}, 126(3):e06625.

\bibitem[{Eymet} et~al., 2009]{Eyme09}
{Eymet}, V., {Fournier}, R., {Dufresne}, J.-L., {Lebonnois}, S., {Hourdin}, F.,
  and {Bullock}, M.~A. (2009).
\newblock {Net exchange parameterization of thermal infrared radiative transfer
  in Venus' atmosphere}.
\newblock {\em J. of Geophys. Res. (Planets)}, 114:E11008.

\bibitem[{Fedorova} et~al., 2016]{Fedo16}
{Fedorova}, A., {Marcq}, E., {Luginin}, M., {Korablev}, O., {Bertaux}, J.~L.,
  and {Montmessin}, F. (2016).
\newblock {Variations of water vapor and cloud top altitude in the Venus'
  mesosphere from SPICAV/VEx observations}.
\newblock {\em Icarus}, 275:143--162.

\bibitem[{Gao} et~al., 2014]{Gao14}
{Gao}, P., {Zhang}, X., {Crisp}, D., {Bardeen}, C.~G., and {Yung}, Y.~L.
  (2014).
\newblock {Bimodal distribution of sulfuric acid aerosols in the upper haze of
  Venus}.
\newblock {\em Icarus}, 231:83--98.

\bibitem[{Garate-Lopez} and {Lebonnois}, 2018]{Gara18}
{Garate-Lopez}, I. and {Lebonnois}, S. (2018).
\newblock {Latitudinal variation of clouds' structure responsible for Venus'
  cold collar}.
\newblock {\em Icarus}, 314:1--11.

\bibitem[{Haus} et~al., 2014]{Haus14}
{Haus}, R., {Kappel}, D., and {Arnold}, G. (2014).
\newblock {Atmospheric thermal structure and cloud features in the southern
  hemisphere of Venus as retrieved from VIRTIS/VEX radiation measurements}.
\newblock {\em Icarus}, 232:232--248.

\bibitem[{Haus} et~al., 2015]{Haus15}
{Haus}, R., {Kappel}, D., and {Arnold}, G. (2015).
\newblock {Radiative heating and cooling in the middle and lower atmosphere of
  Venus and responses to atmospheric and spectroscopic parameter variations}.
\newblock {\em Planetary and Space Science}, 117:262--294.

\bibitem[{Hinson} and {Jenkins}, 1995]{Hins95}
{Hinson}, D.~P. and {Jenkins}, J.~M. (1995).
\newblock {Magellan radio occultation measurements of atmospheric waves on
  Venus}.
\newblock {\em Icarus}, 114:310--327.

\bibitem[{Imamura} et~al., 2017]{Imam17}
{Imamura}, T., {Ando}, H., {Tellmann}, S., {P{\"a}tzold}, M., {H{\"a}usler},
  B., {Yamazaki}, A., {Sato}, T.~M., {Noguchi}, K., {Futaana}, Y.,
  {Oschlisniok}, J., {Limaye}, S., {Choudhary}, R.~K., {Murata}, Y.,
  {Takeuchi}, H., {Hirose}, C., {Ichikawa}, T., {Toda}, T., {Tomiki}, A.,
  {Abe}, T., {Yamamoto}, Z.-i., {Noda}, H., {Iwata}, T., {Murakami}, S.-y.,
  {Satoh}, T., {Fukuhara}, T., {Ogohara}, K., {Sugiyama}, K.-i., {Kashimura},
  H., {Ohtsuki}, S., {Takagi}, S., {Yamamoto}, Y., {Hirata}, N., {Hashimoto},
  G.~L., {Yamada}, M., {Suzuki}, M., {Ishii}, N., {Hayashiyama}, T., {Lee},
  Y.~J., and {Nakamura}, M. (2017).
\newblock {Initial performance of the radio occultation experiment in the Venus
  orbiter mission Akatsuki}.
\newblock {\em Earth, Planets, and Space}, 69:137.

\bibitem[{Imamura} and {Hashimoto}, 2001]{Imam01}
{Imamura}, T. and {Hashimoto}, G.~L. (2001).
\newblock {Microphysics of Venusian Clouds in Rising Tropical Air.}
\newblock {\em Journal of Atmospheric Sciences}, 58(23):3597--3612.

\bibitem[{Imamura} et~al., 2018]{Imam18}
{Imamura}, T., {Miyamoto}, M., {Ando}, H., {H{\"a}usler}, B., {P{\"a}tzold},
  M., {Tellmann}, S., {Tsuda}, T., {Aoyama}, Y., {Murata}, Y., {Takeuchi}, H.,
  {Yamazaki}, A., {Toda}, T., and {Tomiki}, A. (2018).
\newblock {Fine Vertical Structures at the Cloud Heights of Venus Revealed by
  Radio Holographic Analysis of Venus Express and Akatsuki Radio Occultation
  Data}.
\newblock {\em Journal of Geophysical Research (Planets)}, 123(8):2151--2161.

\bibitem[{Jessup} et~al., 2020]{Jess20}
{Jessup}, K.-L., {Marcq}, E., {Bertaux}, J.-L., {Mills}, F.~P., {Limaye}, S.,
  and {Roman}, A. (2020).
\newblock {On Venus' cloud top chemistry, convective activity and topography: A
  perspective from HST}.
\newblock {\em Icarus}, 335:113372.

\bibitem[{Jessup} et~al., 2015]{Jess15}
{Jessup}, K.~L., {Marcq}, E., {Mills}, F., {Mahieux}, A., {Limaye}, S.,
  {Wilson}, C., {Allen}, M., {Bertaux}, J.-L., {Markiewicz}, W., {Roman}, T.,
  {Vandaele}, A.-C., {Wilquet}, V., and {Yung}, Y. (2015).
\newblock {Coordinated Hubble Space Telescope and Venus Express Observations of
  Venus' upper cloud deck}.
\newblock {\em Icarus}, 258:309--336.

\bibitem[{Kerzhanovich} et~al., 1986]{Kerz86}
{Kerzhanovich}, V.~V., {Aleksandrov}, Y.~N., {Andreev}, R.~A., {Armand}, N.~A.,
  {Bakitko}, R.~V., {Blamont}, J.~E., {Boloh}, L., {Hildebrand}, C.~E.,
  {Ignatov}, S.~P., {Ingersoll}, A.~P., {Lysov}, V.~P., {Mottsulev}, B.~I.,
  {Petit}, G., {Pichkhadze}, K.~M., {Preston}, R.~A., {Vorontsov}, V.~A.,
  {Vyshlov}, A.~S., {Young}, R.~E., and {Zaitsev}, A.~L. (1986).
\newblock {Smallscale Turbulence in the Venus Middle Cloud Layer}.
\newblock {\em Soviet Astronomy Letters}, 12:20--22.

\bibitem[{Kouyama} et~al., 2017]{Kouy17}
{Kouyama}, T., {Imamura}, T., {Taguchi}, M., {Fukuhara}, T., {Sato}, T.~M.,
  {Yamazaki}, A., {Futaguchi}, M., {Murakami}, S., {Hashimoto}, G.~L., {Ueno},
  M., {Iwagami}, N., {Takagi}, S., {Takagi}, M., {Ogohara}, K., {Kashimura},
  H., {Horinouchi}, T., {Sato}, N., {Yamada}, M., {Yamamoto}, Y., {Ohtsuki},
  S., {Sugiyama}, K., {Ando}, H., {Takamura}, M., {Yamada}, T., {Satoh}, T.,
  and {Nakamura}, M. (2017).
\newblock {Topographical and Local Time Dependence of Large Stationary Gravity
  Waves Observed at the Cloud Top of Venus}.
\newblock {\em Geophysical Research Letters}, 44:12098--12105.

\bibitem[{Krasnopolsky}, 2007]{Kras07}
{Krasnopolsky}, V.~A. (2007).
\newblock {Chemical kinetic model for the lower atmosphere of Venus}.
\newblock {\em Icarus}, 191(1):25--37.

\bibitem[{Krasnopolsky}, 2012]{Kras12}
{Krasnopolsky}, V.~A. (2012).
\newblock {A photochemical model for the Venus atmosphere at 47-112 km}.
\newblock {\em Icarus}, 218(1):230--246.

\bibitem[{Krasnopolsky}, 2013]{Kras13}
{Krasnopolsky}, V.~A. (2013).
\newblock {S$_{3}$ and S$_{4}$ abundances and improved chemical kinetic model
  for the lower atmosphere of Venus}.
\newblock {\em Icarus}, 225(1):570--580.

\bibitem[{Lebonnois} et~al., 2015]{Lebo15}
{Lebonnois}, S., {Eymet}, V., {Lee}, C., and {Vatant d'Ollone}, J. (2015).
\newblock {Analysis of the radiative budget of the Venusian atmosphere based on
  infrared Net Exchange Rate formalism}.
\newblock {\em J. of Geophys. Res. (Planets)}, 120:1186--1200.

\bibitem[{Lebonnois} et~al., 2010]{Lebo10}
{Lebonnois}, S., {Hourdin}, F., {Eymet}, V., {Crespin}, A., {Fournier}, R., and
  {Forget}, F. (2010).
\newblock {Superrotation of Venus' atmosphere analyzed with a full general
  circulation model}.
\newblock {\em J. of Geophys. Res. (Planets)}, 115:E06006.

\bibitem[{Lebonnois} and {Schubert}, 2017]{Lebo17}
{Lebonnois}, S. and {Schubert}, G. (2017).
\newblock {The deep atmosphere of Venus and the possible role of density-driven
  separation of CO$_{2}$ and N$_{2}$}.
\newblock {\em Nature Geoscience}, pages 473--477.

\bibitem[{Lebonnois} et~al., 2016]{Lebo16}
{Lebonnois}, S., {Sugimoto}, N., and {Gilli}, G. (2016).
\newblock {Wave analysis in the atmosphere of Venus below 100-km altitude,
  simulated by the LMD Venus GCM}.
\newblock {\em Icarus}, 278:38--51.

\bibitem[{Lee} et~al., 2019]{Lee19}
{Lee}, Y.~J., {Jessup}, K.-L., {Perez-Hoyos}, S., {Titov}, D.~V., {Lebonnois},
  S., {Peralta}, J., {Horinouchi}, T., {Imamura}, T., {Limaye}, S., {Marcq},
  E., {Takagi}, M., {Yamazaki}, A., {Yamada}, M., {Watanabe}, S., {Murakami},
  S.-y., {Ogohara}, K., {McClintock}, W.~M., {Holsclaw}, G., and {Roman}, A.
  (2019).
\newblock {Long-term Variations of Venus{\textquoteright}s 365 nm Albedo
  Observed by Venus Express, Akatsuki, MESSENGER, and the Hubble Space
  Telescope}.
\newblock {\em The Astronomical Journal,}, 158(3):126.

\bibitem[{Lef{\`e}vre} et~al., 2018]{Lefe18}
{Lef{\`e}vre}, M., {Lebonnois}, S., and {Spiga}, A. (2018).
\newblock {Three-Dimensional Turbulence-Resolving Modeling of the Venusian
  Cloud Layer and Induced Gravity Waves: Inclusion of Complete Radiative
  Transfer and Wind Shear}.
\newblock {\em Journal of Geophysical Research (Planets)}, 123:2773--2789.

\bibitem[{Lef{\`e}vre} et~al., 2017]{Lefe17}
{Lef{\`e}vre}, M., {Spiga}, A., and {Lebonnois}, S. (2017).
\newblock {Three-dimensional turbulence-resolving modeling of the Venusian
  cloud layer and induced gravity waves}.
\newblock {\em Journal of Geophysical Research (Planets)}, 122:134--149.

\bibitem[{Lef{\`e}vre} et~al., 2020]{Lefe20}
{Lef{\`e}vre}, M., {Spiga}, A., and {Lebonnois}, S. (2020).
\newblock {Mesoscale modeling of Venus' bow-shape waves}.
\newblock {\em Icarus}, 335:113376.

\bibitem[{Lef{\`e}vre} et~al., 2021]{Lefe21}
{Lef{\`e}vre}, M., {Turbet}, M., and {Pierrehumbert}, R. (2021).
\newblock {3D Convection-resolving Model of Temperate, Tidally Locked
  Exoplanets}.
\newblock {\em The Astrophysical Journal}, 913(2):101.

\bibitem[{Lindzen}, 1971]{Lind71}
{Lindzen}, R.~S. (1971).
\newblock {Tides and Gravity Waves in the Upper Atmosphere}.
\newblock In {Fiocco}, G., editor, {\em Mesospheric Models and Related
  Experiments}, volume~25 of {\em Astrophysics and Space Science Library}, page
  122.

\bibitem[{Linkin} et~al., 1986]{Link86a}
{Linkin}, V.~M., {Kerzhanovich}, V.~V., {Lipatov}, A.~N., {Pichkadze}, K.~M.,
  {Shurupov}, A.~A., {Terterashvili}, A.~V., {Ingersoll}, A.~P., {Crisp}, D.,
  {Grossman}, A.~W., {Young}, R.~E., {Seiff}, A., {Ragent}, B., {Blamont},
  J.~E., {Elson}, L.~S., and {Preston}, R.~A. (1986).
\newblock {VEGA balloon dynamics and vertical winds in the Venus middle cloud
  region}.
\newblock {\em Science}, 231:1417--1419.

\bibitem[{Lorenz} et~al., 2018]{Lore18}
{Lorenz}, R.~D., {Crisp}, D., and {Huber}, L. (2018).
\newblock {Venus atmospheric structure and dynamics from the VEGA lander and
  balloons: New results and PDS archive}.
\newblock {\em Icarus}, 305:277--283.

\bibitem[{Mahieux} et~al., 2015]{Mahi15}
{Mahieux}, A., {Vandaele}, A.~C., {Robert}, S., {Wilquet}, V., {Drummond}, R.,
  {Chamberlain}, S., {Belyaev}, D., and {Bertaux}, J.~L. (2015).
\newblock {Venus mesospheric sulfur dioxide measurement retrieved from SOIR on
  board Venus Express}.
\newblock {\em Planetary and Space Science}, 113:193--204.

\bibitem[{Mahieux} et~al., 2021]{Mahi21}
{Mahieux}, A., {Yelle}, R.~V., {Yoshida}, N., {Robert}, S., {Piccialli}, A.,
  {Nakagawa}, H., {Kasaba}, Y., {Mills}, F.~P., and {Vandaele}, A.~C. (2021).
\newblock {Determination of the Venus eddy diffusion profile from CO and
  CO$_{2}$ profiles using SOIR/Venus Express observations}.
\newblock {\em Icarus}, 361:114388.

\bibitem[{Marcq} et~al., 2013]{Marc13}
{Marcq}, E., {Bertaux}, J.-L., {Montmessin}, F., and {Belyaev}, D. (2013).
\newblock {Variations of sulphur dioxide at the cloud top of Venus's dynamic
  atmosphere}.
\newblock {\em Nature Geoscience}, 6:25--28.

\bibitem[{Marcq} et~al., 2020]{Marc20}
{Marcq}, E., {Lea Jessup}, K., {Baggio}, L., {Encrenaz}, T., {Lee}, Y.~J.,
  {Montmessin}, F., {Belyaev}, D., {Korablev}, O., and {Bertaux}, J.-L. (2020).
\newblock {Climatology of SO$_{2}$ and UV absorber at Venus' cloud top from
  SPICAV-UV nadir dataset}.
\newblock {\em Icarus}, 335:113368.

\bibitem[{Markiewicz} et~al., 2007]{Mark07}
{Markiewicz}, W.~J., {Titov}, D.~V., {Limaye}, S.~S., {Keller}, H.~U.,
  {Ignatiev}, N., {Jaumann}, R., {Thomas}, N., {Michalik}, H., {Moissl}, R.,
  and {Russo}, P. (2007).
\newblock {Morphology and dynamics of the upper cloud layer of Venus}.
\newblock {\em Nature}, 450:633--636.

\bibitem[{McGouldrick} and {Toon}, 2007]{Mcgo07}
{McGouldrick}, K. and {Toon}, O.~B. (2007).
\newblock {An investigation of possible causes of the holes in the
  condensational Venus cloud using a microphysical cloud model with a
  radiative-dynamical feedback}.
\newblock {\em Icarus}, 191(1):1--24.

\bibitem[{McGouldrick} and {Toon}, 2008]{Mcgo08}
{McGouldrick}, K. and {Toon}, O.~B. (2008).
\newblock {Observable effects of convection and gravity waves on the Venus
  condensational cloud}.
\newblock {\em Plan. and Sp. Sci.}, 56:1112--1131.

\bibitem[Moeng et~al., 2007]{Moen07}
Moeng, C., Dudhia, J., Klemp, J., and Sullivan, P. (2007).
\newblock {Examining Two-Way Grid Nesting for Large Eddy Simulation of the PBL
  Using the WRF Model}.
\newblock {\em Monthly Weather Review}, 135(6):2295--2311.

\bibitem[{Morellina} and {Bellan}, 2022]{More22}
{Morellina}, S. and {Bellan}, J. (2022).
\newblock {Turbulent chemical-species mixing in the Venus lower atmosphere at
  different altitudes: a direct numerical simulation study relevant to
  understanding species spatial distribution}.
\newblock {\em Icarus}, 371:114686.

\bibitem[{Mori} et~al., 2021]{Mori21}
{Mori}, R., {Imamura}, T., {Ando}, H., {H{\"a}usler}, B., {P{\"a}tzold}, M.,
  and {Tellmann}, S. (2021).
\newblock {Gravity Wave Packets in the Venusian Atmosphere Observed by Radio
  Occultation Experiments: Comparison With Saturation Theory}.
\newblock {\em Journal of Geophysical Research (Planets)}, 126(9):e06912.

\bibitem[{Oschlisniok} et~al., 2021]{Osch21}
{Oschlisniok}, J., {H{\"a}usler}, B., {P{\"a}tzold}, M., {Tellmann}, S.,
  {Bird}, M.~K., {Peter}, K., and {Andert}, T.~P. (2021).
\newblock {Sulfuric acid vapor and sulfur dioxide in the atmosphere of Venus as
  observed by the Venus Express radio science experiment VeRa}.
\newblock {\em Icarus}, 362:114405.

\bibitem[{Parkinson} et~al., 2015]{Park15}
{Parkinson}, C.~D., {Gao}, P., {Esposito}, L., {Yung}, Y., {Bougher}, S., and
  {Hirtzig}, M. (2015).
\newblock {Photochemical control of the distribution of Venusian water}.
\newblock {\em Planetary and Space Science}, 113:226--236.

\bibitem[{Peralta} et~al., 2008]{Pera08}
{Peralta}, J., {Hueso}, R., {S{\'a}nchez-Lavega}, A., {Piccioni}, G.,
  {Lanciano}, O., and {Drossart}, P. (2008).
\newblock {Characterization of mesoscale gravity waves in the upper and lower
  clouds of Venus from VEX-VIRTIS images}.
\newblock {\em J. of Geophys. Res. (Planets)}, 113:E00B18.

\bibitem[{Peralta} et~al., 2019]{Pera19}
{Peralta}, J., {S{\'a}nchez-Lavega}, A., {Horinouchi}, T., {McGouldrick}, K.,
  {Garate-Lopez}, I., {Young}, E.~F., {Bullock}, M.~A., {Lee}, Y.~J.,
  {Imamura}, T., {Satoh}, T., and {Limaye}, S.~S. (2019).
\newblock {New cloud morphologies discovered on the Venus's night during
  Akatsuki}.
\newblock {\em Icarus}, 333:177--182.

\bibitem[{Piccialli} et~al., 2014]{Picc14}
{Piccialli}, A., {Titov}, D.~V., {S\'anchez-Lavega}, A., {Peralta}, J.,
  {Shalygina}, O., {Markiewicz}, W.~J., and {Svedhem}, H. (2014).
\newblock {High latitude gravity waves at the Venus cloud tops as observed by
  the Venus Monitoring Camera on board Venus Express}.
\newblock {\em Icarus}, 227:94--111.

\bibitem[{Rimmer} et~al., 2021]{Rimm21}
{Rimmer}, P.~B., {Jordan}, S., {Constantinou}, T., {Woitke}, P., {Shorttle},
  O., {Hobbs}, R., and {Paschodimas}, A. (2021).
\newblock {Hydroxide Salts in the Clouds of Venus: Their Effect on the Sulfur
  Cycle and Cloud Droplet pH}.
\newblock {\em The Planetary Science Journal}, 2(4):133.

\bibitem[{Rio} et~al., 2010]{Rio10}
{Rio}, C., {Hourdin}, F., {Couvreux}, F., and {Jam}, A. (2010).
\newblock {Resolved Versus Parametrized Boundary-Layer Plumes. Part II:
  Continuous Formulations of Mixing Rates for Mass-Flux Schemes}.
\newblock {\em Boundary-Layer Meteorology}, 135(3):469--483.

\bibitem[{Rossow} et~al., 1980]{Ross80}
{Rossow}, W.~B., {del Genio}, A.~D., {Limaye}, S.~S., and {Travis}, L.~D.
  (1980).
\newblock {Cloud morphology and motions from Pioneer Venus images}.
\newblock {\em Journal of Geophysical Research}, 85:8107--8128.

\bibitem[{Sagdeev} et~al., 1986]{Sagd86}
{Sagdeev}, R.~Z., {Linkin}, V.~M., {Kerzhanovich}, V.~V., {Lipatov}, A.~N.,
  {Shurupov}, A.~A., {Blamont}, J.~E., {Crisp}, D., {Ingersoll}, A.~P.,
  {Elson}, L.~S., {Preston}, R.~A., {Hildebrand}, C.~E., {Ragent}, B., {Seiff},
  A., {Young}, R.~E., {Petit}, G., {Boloh}, L., {Alexandrov}, Y.~N., {Armand},
  N.~A., {Bakitko}, R.~V., and {Selivanov}, A.~S. (1986).
\newblock {Overview of VEGA Venus balloon in situ meteorological measurements}.
\newblock {\em Science}, 231:1411--1414.

\bibitem[{Seiff} et~al., 1980]{Seif80}
{Seiff}, A., {Kirk}, D.~B., {Young}, R.~E., {Blanchard}, R.~C., {Findlay},
  J.~T., {Kelly}, G.~M., and {Sommer}, S.~C. (1980).
\newblock {Measurements of thermal structure and thermal contrasts in the
  atmosphere of Venus and related dynamical observations - Results from the
  four Pioneer Venus probes}.
\newblock {\em Journal of Geophysical Research}, 85:7903--7933.

\bibitem[{Shao} et~al., 2020]{Shao20}
{Shao}, W.~D., {Zhang}, X., {Bierson}, C.~J., and {Encrenaz}, T. (2020).
\newblock {Revisiting the Sulfur-Water Chemical System in the Middle Atmosphere
  of Venus}.
\newblock {\em Journal of Geophysical Research (Planets)}, 125(8):e06195.

\bibitem[{Skamarock} and {Klemp}, 2008]{Skam08}
{Skamarock}, W.~C. and {Klemp}, J.~B. (2008).
\newblock {A time-split nonhydrostatic atmospheric model for weather research
  and forecasting applications}.
\newblock {\em Journal of Computational Physics}, 227:3465--3485.

\bibitem[Spiga et~al., 2010]{Spig10}
Spiga, A., Forget, F., Lewis, S.~R., and Hinson, D.~P. (2010).
\newblock Structure and dynamics of the convective boundary layer on mars as
  inferred from large-eddy simulations and remote-sensing measurements.
\newblock {\em Quarterly Journal of the Royal Meteorological Society},
  136:414--428.

\bibitem[{Tellmann} et~al., 2009]{Tell09}
{Tellmann}, S., {Haeusler}, B., {Paetzold}, M., {Bird}, M.~K., {Tyler}, G.~L.,
  {Andert}, T., and {Remus}, S. (2009).
\newblock {The Structure of the Venus Neutral Atmosphere as seen by the Radio
  Science Experiment VeRa on Venus Express}.
\newblock {\em J. of Geophys. Res. (Planets)}, 114:E00B36.

\bibitem[{Tellmann} et~al., 2012]{Tell12}
{Tellmann}, S., {H{\"a}usler}, B., {Hinson}, D.~P., {Tyler}, G.~L., {Andert},
  T.~P., {Bird}, M.~K., {Imamura}, T., {P{\"a}tzold}, M., and {Remus}, S.
  (2012).
\newblock {Small-scale temperature fluctuations seen by the VeRa Radio Science
  Experiment on Venus Express}.
\newblock {\em Icarus}, 221:471--480.

\bibitem[{Titov} et~al., 2012]{Tito12}
{Titov}, D.~V., {Markiewicz}, W.~J., {Ignatiev}, N.~I., {Song}, L., {Limaye},
  S.~S., {S\'anchez-Lavega}, A., {Hesemann}, J., {Almeida}, M., {Roatsch}, T.,
  {Matz}, K.-D., {Scholten}, F., {Crisp}, D., {Esposito}, L.~W., {Hviid},
  S.~F., {Jaumann}, R., {Keller}, H.~U., and {Moissl}, R. (2012).
\newblock {Morphology of the cloud tops as observed by the Venus Express
  Monitoring Camera}.
\newblock {\em Icarus}, 217:682--701.

\bibitem[{Vandaele} et~al., 2017a]{Vand17a}
{Vandaele}, A.~C., {Korablev}, O., {Belyaev}, D., {Chamberlain}, S.,
  {Evdokimova}, D., {Encrenaz}, T., {Esposito}, L., {Jessup}, K.~L.,
  {Lef{\`e}vre}, F., {Limaye}, S., {Mahieux}, A., {Marcq}, E., {Mills}, F.~P.,
  {Montmessin}, F., {Parkinson}, C.~D., {Robert}, S., {Roman}, T., {Sandor},
  B., {Stolzenbach}, A., {Wilson}, C., and {Wilquet}, V. (2017a).
\newblock {Sulfur dioxide in the Venus atmosphere: I. Vertical distribution and
  variability}.
\newblock {\em Icarus}, 295:16--33.

\bibitem[{Vandaele} et~al., 2017b]{Vand17b}
{Vandaele}, A.~C., {Korablev}, O., {Belyaev}, D., {Chamberlain}, S.,
  {Evdokimova}, D., {Encrenaz}, T., {Esposito}, L., {Jessup}, K.~L.,
  {Lef{\`e}vre}, F., {Limaye}, S., {Mahieux}, A., {Marcq}, E., {Mills}, F.~P.,
  {Montmessin}, F., {Parkinson}, C.~D., {Robert}, S., {Roman}, T., {Sandor},
  B., {Stolzenbach}, A., {Wilson}, C., and {Wilquet}, V. (2017b).
\newblock {Sulfur dioxide in the Venus Atmosphere: II. Spatial and temporal
  variability}.
\newblock {\em Icarus}, 295:1--15.

\bibitem[{Woo} et~al., 1982]{Woo82}
{Woo}, R., {Armstrong}, J.~W., and {Kliore}, A.~J. (1982).
\newblock {Small-scale turbulence in the atmosphere of Venus}.
\newblock {\em Icarus}, 52(2):335--345.

\bibitem[{Woo} and {Ishimaru}, 1981]{Woo81}
{Woo}, R. and {Ishimaru}, A. (1981).
\newblock {Eddy diffusion coefficient for the atmosphere of Venus from radio
  scintillation measurements}.
\newblock {\em Nature}, 289:383.

\bibitem[{Yung} et~al., 2009]{Yung09}
{Yung}, Y.~L., {Liang}, M.~C., {Jiang}, X., {Shia}, R.~L., {Lee}, C.,
  {B{\'e}zard}, B., and {Marcq}, E. (2009).
\newblock {Evidence for carbonyl sulfide (OCS) conversion to CO in the lower
  atmosphere of Venus}.
\newblock {\em Journal of Geophysical Research (Planets)}, 114(16):E00B34.

\bibitem[{Zhang} et~al., 2012]{Zhan12}
{Zhang}, X., {Liang}, M.~C., {Mills}, F.~P., {Belyaev}, D.~A., and {Yung},
  Y.~L. (2012).
\newblock {Sulfur chemistry in the middle atmosphere of Venus}.
\newblock {\em Icarus}, 217(2):714--739.

\end{thebibliography}
\end{document}